\RequirePackage{fix-cm}
\documentclass[twocolumn,epjc3]{svjour3}  
\smartqed  

\usepackage{siunitx}
\usepackage{amsfonts}
\usepackage{amssymb}
\usepackage{amsmath}
\usepackage[T1]{fontenc}
\usepackage{cite}
\usepackage{cancel}
\RequirePackage{graphicx}
\usepackage{dcolumn}
\usepackage{bm}
\usepackage{tablefootnote}
\usepackage{slashed}
\usepackage{breqn}
\usepackage{lineno}
\usepackage{IEEEtrantools}
\usepackage{csquotes}
\usepackage{makecell}
\usepackage{multirow}
\usepackage{cancel}
\usepackage{verbatim}
\usepackage[compat=1.1.0]{tikz-feynman}
\usepackage{siunitx}
\usepackage{booktabs}
\usepackage{graphicx} 
\usepackage[section]{placeins}
\usepackage{needspace}
\usepackage[compatibility=false]{caption}
\usepackage{subcaption}
\usepackage{cuted}
\usepackage{float}
\usepackage{placeins} 
\usepackage[colorlinks = true,
            linkcolor = blue,
            urlcolor  = blue,
            citecolor = blue,
            anchorcolor = blue]{hyperref}
\usepackage{tikz,xcolor}
\usepackage{xspace}


\newcommand{\figref}[1]{Fig.~\ref{#1}}

\newcommand{\tabref}[1]{Tab.~\ref{#1}}
\newcommand{\rinv}{\ensuremath{r_{\text{inv}}}\xspace}

\DeclareOldFontCommand{\tt}{\normalfont\ttfamily}{\mathtt}
\definecolor{lime}{HTML}{A6CE39}
\DeclareRobustCommand{\orcidicon}{
	\begin{tikzpicture}
	\draw[lime, fill=lime] (0,0) 
	circle [radius=0.16] 
	node[white] {{\fontfamily{qag}\selectfont \tiny ID}};
	\draw[white, fill=white] (-0.0725,0.095) 
	circle [radius=0.007];
	\end{tikzpicture}
}

\foreach \x in {A, ..., Z}{\expandafter\xdef\csname orcid\x\endcsname{\noexpand\href{https://orcid.org/\csname orcidauthor\x\endcsname}
			{\noexpand\orcidicon}}
}

\journalname{Noname}
\begin{document} 

\title{Probing the Higgs Portal to a Strongly-Interacting \\ Dark Sector at the FCC-ee}


\author{%
Cesare~Cazzaniga,\thanksref{e1,addr1}\orcidB{}
  Annapaola~de~Cosa,\thanksref{e2,addr1}\orcidC{}
  Felix~Kahlhoefer,\thanksref{e3,addr2}\orcidD{}
  Andrea~S.~Maria,\thanksref{e4,addr1}~\orcidE{}
  Roberto~Seidita,\thanksref{e5,addr1}\orcidF{}
  Emre~Sitti\thanksref{e6,addr0,addr01,addr1}\orcidA{}
}%

\thankstext{e1}{e-mail: cesare.cazzaniga@cern.ch}
\thankstext{e2}{e-mail: adecosa@phys.ethz.ch}
\thankstext{e3}{e-mail: kahlhoefer@kit.edu}
\thankstext{e4}{e-mail: andmaria@ethz.ch}
\thankstext{e5}{e-mail: roberto.seidita@cern.ch}
\thankstext{e6}{e-mail: emre.sitti@uzh.ch} 

\institute{
University of Zurich, Faculty of Science,  CH-8057 Zurich, Switzerland 
\label{addr0}
\and
\text{ }PSI Center for Neutron and Muon Sciences, 5232 Villigen PSI, Switzerland
\label{addr01}
\and
\text{ }ETH Zurich, Institute for Particle Physics and Astrophysics, CH-8093 Zurich, Switzerland 
\label{addr1}
\and 
\text{ }Institute for Astroparticle Physics (IAP), Karlsruhe Institute of Technology (KIT), Hermann-von-Helmholtz-Platz 1, 76344 Eggenstein-Leopoldshafen, Germany
\label{addr2}
}

\date{Received: date / Accepted: date}

\maketitle

\hfill \parbox{17.0cm}{\vspace{-7.9cm}\flushright P3H-25-077, TTP25-036 \vspace{7cm}}

\vspace{-4mm}

\begin{abstract}
 
This work explores exotic signatures from confining dark sectors that may arise in the $e^+e^-$ collision mode at the Future Circular Collider. Assuming the Higgs boson mediates the interaction between the Standard Model and the dark sector, dark quarks can be produced in $e^+e^-$ collisions. The ensuing strong dynamics may lead to semi-visible jet final states, containing both visible and invisible particles.

 We investigate semi-visible jets with different fractions of invisible states, and enriched in leptons and photons. When the invisible component is large, selections based on kinematic features, such as the missing energy in the event, already provide good signal-to-background discrimination. For smaller invisible fractions, the reduced missing energy makes these signals more similar to Standard Model events, and we therefore employ a graph neural network jet tagger exploiting differences in jet substructure. This machine learning strategy improves sensitivity and enhances the discovery prospects of Higgs boson-induced semi-visible jets at the Future Circular Collider. Our results show that the proposed strategy can effectively probe a wide parameter space for the models considered, and a variety of signatures, constraining the Higgs boson exotic branching ratios into dark quarks at the permille-level.

\end{abstract}

\vfill

\section{Introduction}
\label{intro}

There is strong evidence that the Standard Model (SM) is not enough to explain all known physics phenomena, including the origin of dark matter (DM), neutrino masses and the baryon asymmetry.
Astrophysical and cosmological measurements of the gravitational effects of DM at large scales~\cite{Bertone:2004pz,Planck2018} suggest the existence of a separate sector of particles and interactions, commonly referred to as the dark sector (DS). The lack of experimental confirmation for minimal DS hypotheses suggests that these sectors may be more complex, potentially manifesting through unusual signatures at particle colliders.
The Hidden Valley (HV) scenario~\cite{Strassler_2007} proposes alternative Beyond the SM (BSM) models by introducing a strongly-coupled DS. The visible and dark sectors may interact via different mediators, including the Higgs boson. These models can be compatible with explanations for DM~\cite{Bernreuther_2020,Beauchesne:2018myj} and neutral naturalness~\cite{Craig:2015pha}, and offer unconventional signatures that may have gone unnoticed at particle colliders.

Under the assumption that the DS behaves similarly to QCD in the SM, semi-visible jet (SVJ) signatures~\cite{Cohen:2015toa,Cohen:2017pzm,Beauchesne:2017yhh,Beauchesne:2018myj,Bernreuther_2020,Beauchesne:2019ato,Knapen_2021} arise naturally within the HV scenario. These final states may also include additional photons~\cite{Cazzaniga:2024mmv} and leptons~\cite{Cazzaniga:2022hxl,Beauchesne_2023}, alongside hadronic activity, thereby enriching the range of potential signatures at particle colliders.

As part of the FCC-ee research programme, a dedicated Higgs boson factory phase is planned, with the centre-of-mass energy set slightly above the $ZH$ production threshold~\cite{Azzurri:2021nmy,Ruan:2014xxa,Antusch:2021fcc}. This phase will enable precise measurements of Higgs boson's couplings and provide an opportunity to explore its exotic decay channels. Motivated by this, we investigate the potential role of the Higgs boson as a portal to the HV, specifically leading to SVJ signatures. 
Furthermore, the FCC-ee offers a particularly promising environment for studying these signatures due to its significantly reduced hadronic activity. The absence of background from QCD processes, which often overwhelm potential signals at hadron colliders such as the Large Hadron Collider (LHC), enhances the sensitivity to such exotic phenomena. This cleaner environment allows for more precise identification and analysis of SVJ signatures.

Assuming that the Higgs boson decays to dark quarks, the dark QCD dynamics leads to a shower and hadronization in the DS resulting in a spectrum of dark hadrons. The number of stable and unstable dark hadrons produced in the dark hadronization process can vary depending on the details of the DS. In previous literature~\cite{Cohen:2015toa,Cohen:2017pzm}, an effective invisible fraction parameter has been defined as
\[
\rinv =  \left\langle \frac{N_{\text{stable}}}{N_{\text{stable}} + N_{\text{unstable}}} \right\rangle,
\]
where $N_{\text{stable}}$ and $N_{\text{unstable}}$ are the number of stable and unstable dark hadrons respectively, with the latter decaying back to SM particles. This invisible fraction allows to capture variations in the details of the DS and defines the parameter space of SVJs in terms of the amount of missing energy ($E_{\mathrm{miss}}$).  

To explore these signatures, we construct benchmark models with varying \rinv. The high-\rinv regime, where most dark hadrons are stable, is motivated by the potential connection to the DM puzzle. Here, selections based on event kinematics, leveraging the larger expected amount of $E_{\mathrm{miss}}$ in signal events, already provide good signal--background separation. In contrast, when moving to the low-\rinv regime, the smaller $E_{\mathrm{miss}}$ in signal events does not allow for powerful discrimination against SM processes. In both regimes, further discrimination is achieved by means of a physics-informed jet tagging algorithm based on graph neural networks (GNNs). By exploiting the distinctive differences in the jet substructure between SVJs and SM jets, the GNN restores sensitivity and significantly enhances discovery prospects in the low-\rinv regime. Overall, our results indicate that this combined strategy has the potential to probe a large parameter space, reaching sensitivity to Higgs boson exotic branching ratios to dark quarks at the permille-level.

\section{Model setup}
\label{sec:Model}
 
Driven by our target experimental signature, which primarily consists of SVJs, we construct a HV model tailored to estimate the FCC-ee sensitivity to this class of signals. In addition to hadronic activity, the model allows for the prompt production non-isolated leptons and photons within the jets. The presence of these additional states ensures that the benchmark captures a wide class of theoretically motivated scenarios.

In the HV scenario that we consider, the SM gauge group is supplemented by a non-abelian dark sector $SU(N_c)_d$. The fermions in the fundamental representation of $SU(N_c)_d$ are called the dark quarks $q_{d,i}$, with $i \in \{1, \cdots , N_f \}$, where $N_f$ is the number of dark flavors. Here we choose  the number of dark color charges $N_c =3$ and $N_f=2$ for concreteness. The DS has dynamics similar to SM QCD~\cite{Albouy:2022cin} and is assumed not to be completely secluded from the SM. Multiple portal interactions are introduced to enable the production of dark quarks in $e^+ e^-$ collisions and the subsequent decays of the unstable dark bound states. 

In order to allow for the dark quarks to be produced at the FCC-ee Higgs boson factory phase, we consider the following portal interaction with coupling $y$ mediated by the Higgs boson field $h$:
\begin{equation}
    \mathcal{L}_{h,q_d} \supset - yh \bar{q}_d q_d \; ,\label{eq:HiggsPortal_Lagrangian}
\end{equation}
which induces the decay $H \to q_d \bar{q}_d$. 
The corresponding partial decay width is given by
\begin{equation}\label{eq:HiggsPortal_width}
\Gamma(H\to \bar q_d q_d)
=
\frac{N_c\, y^2\, m_H}{16\pi}
\left(1-\frac{4m_{q_d}^2}{m_H^2}\right)^{3/2},
\end{equation}
where $m_H$ is the Higgs boson mass, and $m_{q_d}$ are the dark quarks masses.


The showering and hadronization of the dark quarks then leads to the formation of dark hadrons of spin 0 (pseudoscalar mesons $\pi_d,\eta_d'$) and spin 1 (vector mesons $\rho_d$) at the dark confinement scale $\Lambda$. Some of these dark hadrons will be stable, and others will decay back to SM according to the conservation of global symmetries of the DS~\cite{Strassler_2007,Cohen:2015toa,Cohen:2017pzm}. 
Motivated by the DM relic abundance~\cite{Bernreuther_2020}, we consider the scenario where the dark pions $\pi^{\pm,0}$ and $\rho_d^{\pm}$ are stable due to dark G-parity~\cite{Bernreuther_2020,Berlin:2018tvf} and dark isospin, respectively. Thus, only the $\rho_d^0$ and ${\eta_d}'$ can decay back into SM particles. We allow the decays of these bound states to SM leptons, quarks and photons via different portal interactions.
The decay of the $\rho_d^0$-meson to SM can be induced by kinetic mixing with the SM photon, such that the $\rho_d^0$-meson couples to SM fermions proportionally to charge. The partial widths of an unstable dark vector meson $\rho_d$ decaying to SM leptons and quarks can then be calculated from a chiral EFT~\cite{Ilten_2018,Knapen_2021}.
The ${\eta_d}'$ can couple to the SM in the same way as an axion-like particle (ALP). Here we focus on the coupling to photons and assume $BR({\eta_d}' \to \gamma \gamma)=100\%$.
For the present study, we assume all dark mesons to decay promptly. As discussed in Refs.~\cite{Knapen_2021,Cazzaniga:2022hxl,Cazzaniga:2024mmv}, this assumption is consistent with experimental constraints on dark hadrons with masses above the GeV-scale.
\begin{figure}[t]
\centering
\begin{tikzpicture}
\begin{feynman}
    \vertex (a) at (-1,1.5) {\( e^+ \)};
    \vertex (b) at (-1,-1.5) {\( e^- \)};
    
    \vertex (c) at (0,0);
    \vertex (d) at (1.25,0);
    
    \vertex (e) at (2,1) ;
    \vertex (f) at (2,-1);
    
    \vertex (g) at (3.25,1.5);
    \vertex (h) at (3.25,0.5);

    \vertex (i) at (3.75,-0.25);
    \vertex (k) at (3.75,-1.75);

    \vertex (j1) at (3,-0.55);
    \vertex (j2) at (3,-1.35);

    \vertex (j3) at (3.62,-0.85);
    \vertex (j4) at (3.62,-1.15);

    \vertex (h1) at (4.3,-0.3); 
    \vertex (h2) at (4.4,-1);   
    \vertex (h3) at (4.3,-1.7); 
    
    \vertex (u1) at (5,0.2) {\( \eta_d' \)};
    \vertex (u2) at (5.5,-1) {\( \rho_d^{\pm}, \pi_d^{0,\pm} \)};
    \vertex (u3) at (5,-2.25) {\( \rho_d^{0}\)};

    \vertex (ph1) at (6,-0.25);
    \vertex (ph2) at (6,0.75);

    \vertex (fer1) at (6,-1.75);
    \vertex (fer2) at (6,-2.75);

    \vertex (hadronization) at (3.75,-1);

    \filldraw[gray!80, thick, draw= black] (4,-1) ellipse (0.4cm and 1cm);

    \diagram* {
        (a) -- [anti fermion] (c) -- [anti fermion] (b),
        (c) -- [boson, edge label=\( Z^* \)] (d),
        (d) -- [scalar, edge label'=\( H \)] (f),
        (d) -- [boson, edge label=\( Z \)] (e),
        (g) -- [fermion, edge label'=\( \ell^+ \)] (e) -- [fermion, edge label'=\( \ell^- \)] (h),
        
        (i) -- [fermion, edge label'=\( \bar{q}_d \)] (f) -- [fermion, edge label'=\( q_d \)] (k),

        (j1) -- [gluon] (j3),
        (j2) -- [gluon] (j4),

        (h1) -- (u1),
        (h2) -- (u2),
        (h3) -- (u3),

        (u1) -- [photon, edge label'=\(\gamma\)] (ph1),
        (u1) -- [photon, edge label=\(\gamma\)] (ph2),

        (u3) -- [fermion, edge label=\(f\)] (fer1),
        (u3) -- [anti fermion, edge label'=\(\bar f\)] (fer2)
    };
\end{feynman}
\end{tikzpicture}

\begin{tikzpicture}
\begin{feynman}
    \vertex (a) at (-1,1.5) {\( e^+ \)};
    \vertex (b) at (-1,-1.5) {\( e^- \)};
    
    \vertex (c) at (0,0);
    \vertex (d) at (1.25,0);
    
    \vertex (e) at (2,1) ;
    \vertex (f) at (2,-1);
    
    \vertex (g) at (3.25,1.5);
    \vertex (h) at (3.25,0.5);

    \vertex (i) at (3.75,-0.25);
    \vertex (k) at (3.75,-1.75);

    \vertex (j1) at (3,-0.55);
    \vertex (j2) at (3,-1.35);

    \vertex (j3) at (3.62,-0.85);
    \vertex (j4) at (3.62,-1.15);

    \vertex (h1) at (4.3,-0.3); 
    \vertex (h2) at (4.4,-1);   
    \vertex (h3) at (4.3,-1.7); 
    
    \vertex (u2) at (5.0,-1){\(\pi_d^{0,\pm} \)};

    \vertex (f2) at (6.5,0.7);
    \vertex (f1) at (6.5,-0.3);

    \vertex (fer) at (6.0,-0.25){\( \bar{f}f \)};
    \vertex (ph) at (6.0,-1){\(\gamma \gamma \)};
    \vertex (inv) at (6.0,-1.75){\( \mathrm{inv} \)};

    \vertex (hadronization) at (3.75,-1);

    \filldraw[gray!80, thick, draw= black] (4,-1) ellipse (0.4cm and 1cm);

    \diagram* {
        (a) -- [anti fermion] (c) -- [anti fermion] (b),
        (c) -- [boson, edge label=\( Z^* \)] (d),
        (d) -- [scalar, edge label'=\( H \)] (f),
        (d) -- [boson, edge label=\( Z \)] (e),
        (g) -- [fermion, edge label'=\( \ell^+ \)] (e) -- [fermion, edge label'=\( \ell^- \)] (h),
        
        (i) -- [fermion, edge label'=\( \bar{q}_d \)] (f) -- [fermion, edge label'=\( q_d \)] (k),

        (j1) -- [gluon] (j3),
        (j2) -- [gluon] (j4),
        (h2) -- (u2),
        (u2) -- (fer),
        (u2) -- (ph),
        (u2) -- (inv),
    };
\end{feynman}
\end{tikzpicture}
\caption{Sketch of the process studied in this work: $e^+e^-\to Z H$ followed by the decay $H\to q_d\bar q_d$ and dark showering and hadronization. The top panel corresponds to the specific model that we use in the high-$r_{\text{inv}}$ regime, where the dark shower contains \(\eta_d'\), \(\rho_d\) and \(\pi_d\) and we consider the decays \(\eta_d'\to\gamma\gamma\) and \(\rho_d^0\to f\bar f\). The bottom panel corresponds to the parametric model that we use in the low-$r_{\text{inv}}$ regime.}
\label{fig:dark_hadronization}
\end{figure}
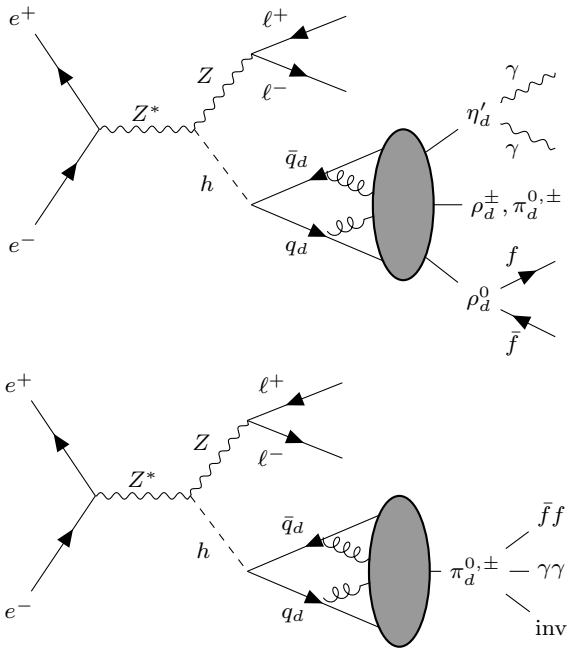

As sketched in the top panel of~\figref{fig:dark_hadronization}, this model leads to SVJs containing leptons and photons produced by the unstable dark hadrons. We will see that this model predicts $\rinv \sim \SIrange{75}{95}{\%}$, depending on the dark shower parameters. To also study models with small \rinv, we additionally introduce a parametric model. This framework is not intended to specify the detailed structure of the DS, but rather to provide a flexible description that can capture a broad class of possible BSM signatures. In this setup, the spectrum is restricted to dark pions $\pi_d^{0,\pm}$, which are allowed to decay into SM leptons, photons, and quarks, while also admitting invisible decays with a branching fraction that parameterizes the invisible fraction. The generality of this model enables us to conduct an investigation of the low-$r_{\text{inv}}$ regime in a model-agnostic manner. The corresponding production mechanism is shown in the bottom panel of \figref{fig:dark_hadronization}.

\section{Monte Carlo Simulations} 
\label{sec:MC_simulation}
All the Monte Carlo samples are generated within the FCC-ee simulation and analysis framework~\cite{FCCAnalyses} using {\tt Pythia8}~\cite{Bierlich:2022pfr}.
The HV module~\cite{Carloni:2011kk,Carloni:2010tw} in {\texttt{Pythia8}} is used to simulate the showering, hadronization, and decays. 
Detector effects are simulated with {\texttt{Delphes3}}~\cite{deFavereau:2013fsa} using the IDEA detector~\cite{IDEAStudyGroup:2025gbt} parameterization.
In our study, anti-$k_T$ jets~\cite{Cacciari:2008gp,Cacciari:2011ma} with \mbox{$R = 1.5$} are reconstructed. We employ large-radius jets to better capture the physics of SVJs, which are expected to be broader than SM jets due to the multi-step shower processes occurring in both the DS and the SM. Additionally, the radiation pattern of SVJs is further broadened by the non-negligible masses of the visibly decaying dark hadrons, which can result in large opening angles between their decay products. Finally, since the Higgs boson is produced approximately at rest, the resulting jets are anticipated to be wider than those typically observed at the LHC.
All samples have been normalized to the cross-section prediction computed with \texttt{Pythia8}.
The signal process $e^{+}e^{-} \to Z(\ell \bar{\ell})H(q_d \bar{q}_d)$ was generated at leading order (LO). We normalize the signal samples according to the experimentally allowed bound for the exotic Higgs boson branching ratio~$\sim 18\%$. The center-of-mass energy was chosen to be $\SI{240}{GeV}$, corresponding to the nominal energy forseen for the ZH run of the FCC-ee, and the total integrated luminosity was set to $10.8~\mathrm{ab^{-1}}$~\cite{FCC:2025lpp}.

For the signal processes in the high $r_{\text{inv}}$ regime, $5 \cdot 10^6$ events have been generated for each benchmark point in~\tabref{tab:benchmark_high}, scanning over the following three parameters: $p_v$ (\texttt{HiddenValley:probVector}), which determines the multiplicity of the dark vector mesons, $p_{\eta}$ \mbox{(\texttt{HiddenValley:\allowbreak probKeepEta1})}, which determines the production rate of $\eta_d'$ meson, and the dark confinement scale $\Lambda$ (\texttt{HiddenValley:Lambda}), which fixes the overall mass scale for the dark bound states. Typically, the $\rho_d^0$ meson mass is expected to be of a similar order of magnitude as $4\pi\Lambda$~\cite{Albouy_2022}. The dark pseudoscalar $\eta_d'$ and vector $\rho_d$ meson masses can in principle differ according to the non-perturbative dynamics of the DS, even for mass-degenerate dark quarks. In our model we assume their masses to be equal. For each choice of $\Lambda$, we fix the masses of the dark mesons by employing the lattice fits approach proposed in Ref.~\cite{Albouy_2022}, and setting $m_{\rho_d^0}/{m_{\pi_d^0}} = 1.75$. This choice ensures that the dark vector mesons cannot decay into dark pions and must decay visibly into SM particles.

\begin{table}[t]
\centering
\scriptsize 
\sffamily
\renewcommand{\arraystretch}{2}
\addtolength{\tabcolsep}{-0.2em}
\begin{tabular}{cccccccc}
\toprule
\textbf{PARAMETER} & \multicolumn{7}{c}{\textbf{BENCHMARK}} \\ \midrule
\(\Lambda~(\mathrm{GeV})\)          & 0.521 & 1.042 & 2.083 & 3.125 & 4.167 & 5.208 & 6.25 \\
\(m_{\pi_d}~(\mathrm{GeV})\)         & 1 & 2 & 4 & 6 & 8 & 10 & 12 \\
\(m_{\rho_d,\eta'_d}~(\mathrm{GeV})\)& 1.75 & 3.5 & 7.0 & 10.5 & 14.0 & 17.5 & 21.0 \\
\midrule
\(p_v\)                              & & 0.1 & & 0.25 & &  0.5 & \\
\midrule
\(p_{\eta}\)                         & 0.25 & & 0.50 & & 0.75 & & 1.00 \\
\bottomrule
\end{tabular}
\caption{Parameter combinations considered as benchmarks for the high-\(r_{\text{inv}}\) regime.}
\label{tab:benchmark_high}
\end{table}

For each set of dark meson masses, we scan over several values of $p_v,~p_{\eta}$ as reported in~\tabref{tab:benchmark_high}. Varying these parameters changes the multiplicity of visibly decaying dark mesons and therefore corresponds to a variation in $r_{\text{inv}}$, see~\ref{app:plot_rinv} for details. However, the average multiplicity of stable (and hence invisible) dark mesons is always larger than the one of decaying dark mesons, and hence $r_{\text{inv}}$ is always well above 0.5. 

To explore the low-$r_{\text{inv}}$ regime,  we adopt a more model-agnostic approach.
We fix $p_v = p_{\eta} = 0$, such that only dark pions are produced in the dark shower and treat $r_{\text{inv}}$ itself as the only free parameter to be scanned together with $\Lambda$. For each signal benchmark point, we generate $5 \times 10^6$ events in the same manner. The benchmark values considered for this low-$r_{\text{inv}}$ case, and the corresponding dark meson masses are summarized in~\tabref{table:signal_parameterization_lowrinv}.

\begin{table}[t]
\centering
\scriptsize 
\sffamily
\renewcommand{\arraystretch}{2}
\addtolength{\tabcolsep}{-0.2em}
\begin{tabular}{cccccccc}
\toprule
\textbf{PARAMETER} & \multicolumn{7}{c}{\textbf{BENCHMARK}} \\ \midrule
\(\Lambda~(\mathrm{GeV})\)  & 0.521 & 1.042 & 2.083 & 3.125 & 4.167 & 5.208 & 6.25 \\
\(m_{\pi_d}~(\mathrm{GeV})\) & 1 & 2 & 4 & 6 & 8 & 10 & 12 \\
\midrule
\(r_{\text{inv}}\) & 0.05 & 0.10 & 0.20 & 0.30 & 0.40 & 0.50 & 0.60 \\
\bottomrule
\end{tabular}
\caption{Parameter combinations considered as benchmarks for the low \(r_{\text{inv}}\) regime.}
\label{table:signal_parameterization_lowrinv}
\end{table}

For the treatment of visible branching fractions into charged leptons, quarks and photons, we distinguish between two cases. In the high-$r_{\text{inv}}$ case, we determine the visible and invisible fractions from the $\rho_d$-mass as predicted by chiral perturbation theory, and propagate them to the corresponding final states.
In the low-$r_{\text{inv}}$ regime we adopt a simplified decay scheme, in which the three different types of particles are produced with equal probability:
\[
\mathrm{BR}_\ell=\mathrm{BR}_q=\mathrm{BR}_\gamma=\tfrac{1-r_{\text{inv}}}{3}.
\]
Within each decay mode, the branching ratios are distributed uniformly over the kinematically accessible flavors.




As background processes we consider $e^{+}e^{-} \to ZZ$, $e^{+}e^{-} \to WW$ and $e^{+}e^{-} \to ZH$. All background samples have been generated at LO with {\tt Pythia8}. The $Z(\ell \ell)H$ events ($5 \cdot 10^6$ simulated events) represent the major background process for the proposed signal due to the identical event topology. The $ZZ$ ($5 \cdot 10^6$ simulated events) process constitutes the second most important background due to its similar final states compared to the signal and comparable cross section. Finally, the $W^{+}W^{-}$ background ($5 \cdot 10^6$ simulated events) is the electroweak background with the highest cross-section and typically should be considered when looking at the $ZH$ channel due to the presence of final state jets and leptons mimicking the same final states as the signal.


\section{Search Strategies}\label{sec:strategy}

To determine the sensitivity of the FCC-ee to the proposed model, we apply a set of kinematic requirements to select the signal topology and reject the background. In our study we focus on the cleanest final state signature for the $ZH$ production mode of SVJs characterized by two leptons from the $Z$ boson and two jets from the Higgs boson.

In particular, we focus on isolated electrons or muons from the $Z$ boson to tag the event. Thus, we select events with at least two  electrons or muons with isolation $\mathrm{I}(\ell)<0.5$, and with opposite charge. The two leptons can be used to reconstruct the mass of the recoiling object $m_{\text{rec}} = \sqrt{s -2E_{\ell\ell}\sqrt{s}+m^2_{\ell\ell}}$, where $E_{\ell\ell}$ and $m_{\ell\ell}$ are the di-lepton system energy and invariant mass, respectively. Since we are interested in events where SVJs are produced directly from the Higgs boson, we require $m_{\text{rec}}$ to be within a window centered around the SM Higgs boson mass ($\SI{120}{GeV} < m_{\text{recoil}} < \SI{130}{GeV}$). Finally, we require the presence of at least two jets with $E>\SI{10}{GeV}$ and $|{\eta}|\leq2.4$. 

\begingroup
\emergencystretch=1.2em\relax 
 In the high-\rinv regime, a $E_{\mathrm{miss}}$ selection is well motivated:
a sizeable invisible component carries away momentum and appears as $E_{\mathrm{miss}}$. We therefore
optimized rectangular $E_{\mathrm{miss}}$ windows together with the minimum azimuthal separation between the $E_{\mathrm{miss}}$ vector
and the jets, $\min\Delta\phi$, and assessed the Asimov
discovery significance $Z_{\mathrm{A}}$~\cite{Cowan_2011} across the scanned parameter space.
\emergencystretch=1.2em\relax
The best overall choice was found to be
$\text{$30<E_{\mathrm{miss}}$}<{90}~{\text{GeV}}$ with $\min\Delta\phi<1.7$. 

\endgroup


The following requirements are applied to select signal-like candidates:

   \begin{enumerate}
    \item Exactly one opposite-sign, isolated lepton pair of the same flavor ($\ell = e, \mu$) with 
          $E > \SI{10}{GeV}$, $|\eta| \leq 2.4$, and $\mathrm{I}(\ell) < 0.5$.
    \item Veto pairs of isolated, different-flavour leptons with 
          $E > \SI{10}{GeV}$, $|\eta| \leq 2.4$, and $\mathrm{I}(\ell) < 0.5$.
    \item At least two jets with 
          $E > \SI{10}{GeV}$ and $|\eta| \leq 2.4$.
    \item $Z$ boson mass window: 
          $85 < m_{\ell\ell} < \SI{95}{GeV}$.
    \item Higgs recoil mass window: 
          $120 < m_{\text{recoil}} < \SI{130}{GeV}$.
    \item Missing energy requirement:
          $30 < E_{\text{miss}} < \SI{90}{GeV}$ (high-$r_{\text{inv}}$).
    \item Azimuthal separation cut:
          $\min\Delta\phi < 1.7$ (high-$r_{\text{inv}}$,~a), 
          $\min\Delta\phi < 0.7$ (low-$r_{\text{inv}}$,~b).
  \end{enumerate}



  \label{tab:selection_cuts}

Each step of the selection targets specific backgrounds while retaining signal events. The full selection efficiencies resulting from~\tabref{tab:selection_cuts} can be found in~\ref{app:sigeff}, in both high and low-\rinv regimes. The di-lepton requirement strongly suppresses the $ZZ$ and $WW$ backgrounds, while the jet and $Z$-mass selections further reduce the $WW$ contribution to below $0.2\%$ of its original rate. 
The Higgs boson recoil mass requirement strongly rejects most of the $ZZ$ and $WW$ backgrounds, while still retaining about one third of the signal and SM $ZH$ background. 
In the high-\rinv regime, the $E_{\mathrm{miss}}$ selection further removes around 80\% of the SM $ZH$ background, while retaining most of the signal.
The resulting cutflow for the high-$r_{\mathrm{inv}}$ case is shown in~\tabref{tab:cutflow_highRinv}. 

\begin{table}[t]
  \centering
  \begin{tabular}{l r r r r}
    \toprule
    \textbf{Step} & $ZZ\,(\%)$ & $ZH\,(\%)$ & $WW\,(\%)$ & Signal\,(\%) \\
    \midrule
    i),ii) & 13 & 53.6 & 2.52 & 55.1 \\
    iii)   & 12.1 & 53.2 & 2.36 & 55.0 \\
    iv)    & 7.62 & 42.5 & 1.51$\times10^{-1}$ & 44.3 \\
    v)     & 2.78$\times10^{-1}$ & 34.9 & 2.2$\times10^{-2}$ & 36.4 \\
    vi)    & 6.5$\times10^{-2}$ & 6.08 & 4.2$\times10^{-4}$ & 31.4 \\
    viia)  & 6.5$\times10^{-2}$ & 5.8 & 3.2$\times10^{-4}$ & 30.7 \\
    \bottomrule
  \end{tabular}

  \caption{Cumulative selection efficiencies for the high-\rinv analysis. 
  The signal efficiencies are averaged over all benchmarks.}
  \label{tab:cutflow_highRinv}
\end{table}

When applying the same baseline selections to the samples in the low-\rinv regime we make two key observations. First, a stringent requirement on the $E_{\mathrm{miss}}$ is not an effective discriminator in this region. Because the invisible energy fraction is intrinsically smaller, a high $E_{\mathrm{miss}}$ threshold removes more of the signal than of the dominant electroweak backgrounds, thereby degrading the overall significance. Second, while the \mbox{$\min\Delta\phi<1.7$} selection cut retains most of the signal in the high-\rinv regime, at low and intermediate \rinv this requirement was found to be too loose. A threshold of $\min\Delta\phi<0.7$ provides a more effective discrimination: It suppresses events in which jets are recoiling against $E_{\mathrm{miss}}$, especially for the $WW$ background. The corresponding cutflow information in this regime can be found in~\tabref{tab:cutflow_lowRinv}.

\begin{table}[t]
  \centering

    \begin{tabular}{l r r r r}
    \toprule
      \textbf{Step} & $ZZ (\%)$& $ZH (\%)$ & $WW (\%)$ & Signal (\%)\\
      \midrule
      i),ii)          & 13 & 53.6 & 2.52 & 55.6 \\
      iii)                    & 12.1 & 53.2 & 2.36 & 54.7 \\
      iv)       & 7.62 & 42.5 & 0.151 & 43.0 \\
      v) & 0.278 & 35.3 &  0.022 & 36.4 \\
      viib)        &  0.168 &  24.4 &  0.005 & 23.8 \\
      \bottomrule
    \end{tabular}

  \caption{Cumulative selection efficiencies for the low-\rinv analysis. 
  For the signal, we have averaged over all samples.}
  \label{tab:cutflow_lowRinv}
\end{table}

\section{SVJ tagging through GNN}
\label{sec:gnn}


While the selections introduced in the previous section are able to strongly suppress the backgrounds from $ZZ$ and $WW$ processes, the $ZH$ background is only slightly suppressed relative to the signal in the high-\rinv case and almost unchanged in the low-\rinv case. To improve signal discrimination, we employ the \texttt{LundNet}~\cite{Dreyer:2020brq} GNN. In this approach, each jet is represented as a Lund graph 
that encodes its full clustering history. Nodes correspond to pseudojets, meaning either single input particles or intermediate clusters produced during the sequential recombination (four-vector sum). Each node is endowed with kinematic and particle-content features, while edges reflect the parent-child relations from clustering, thereby preserving the complete formation history of the jet. Since this evolution is expected to differ between QCD jets and semi-visible dark jets, such information provides a powerful handle for discrimination.
\begin{figure*}[!t]
  \centering
    \includegraphics[width=0.49\linewidth]{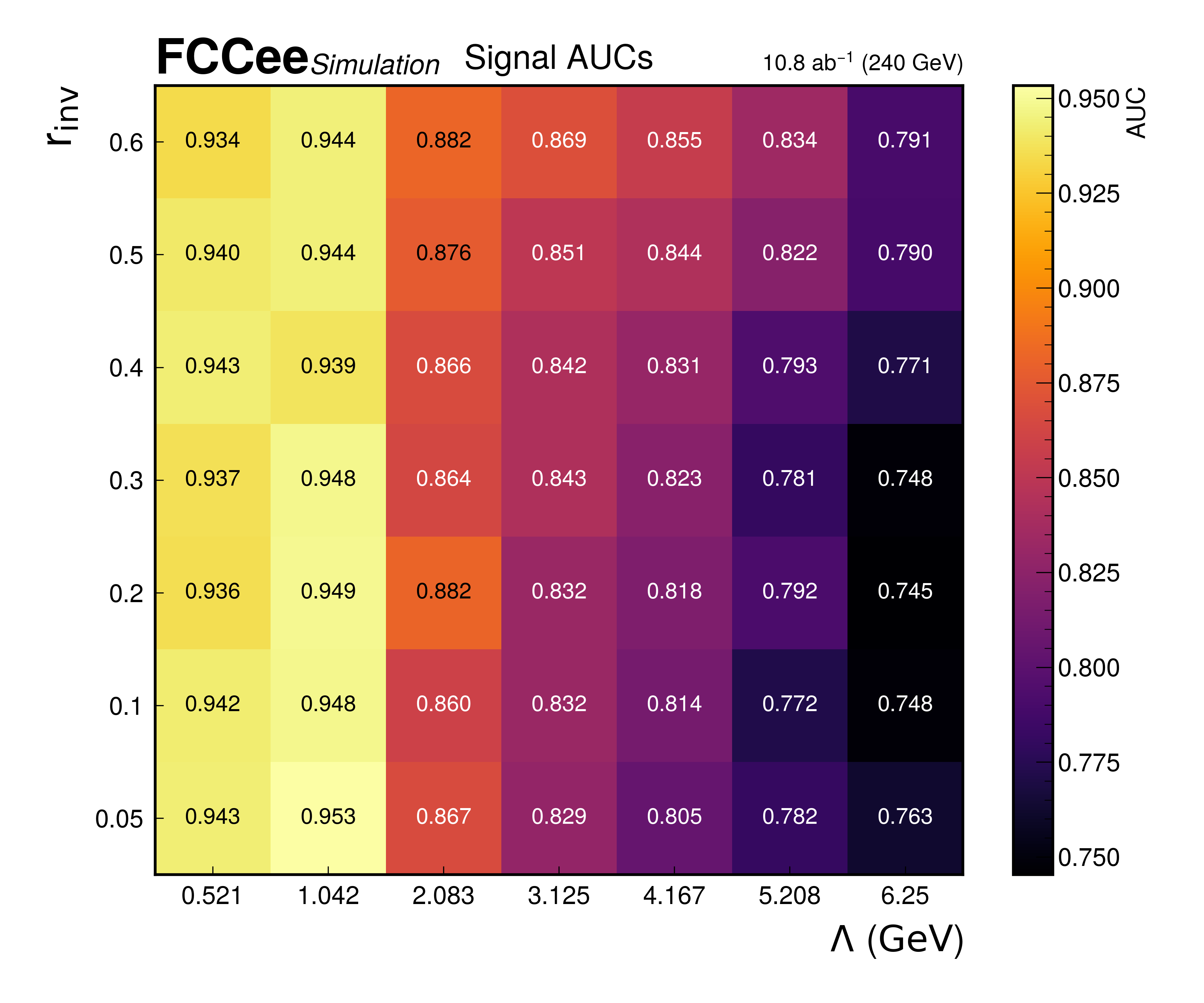}    \includegraphics[width=0.49\linewidth]{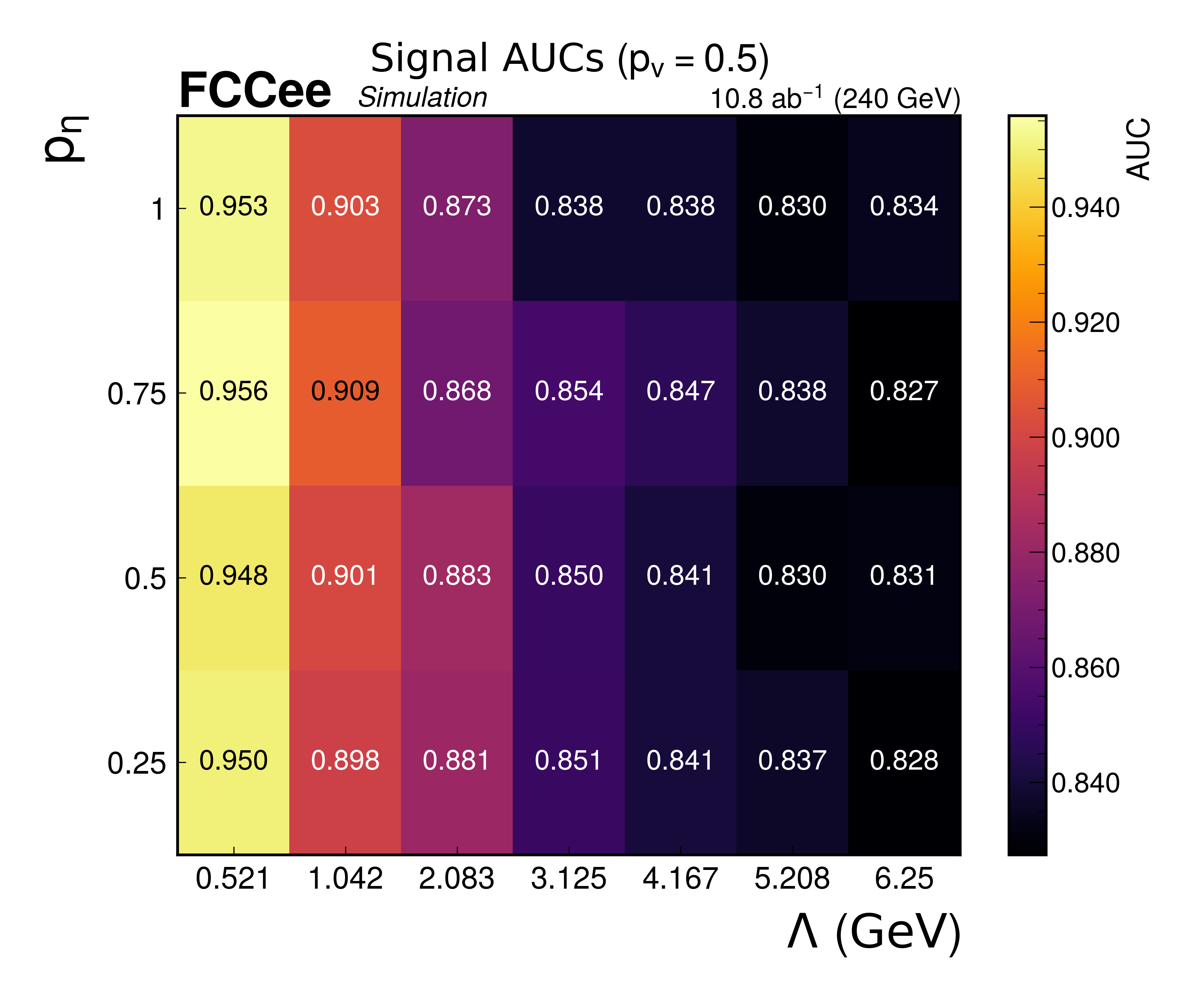}

  \caption{Signal-wise AUC scores in the low-$r_{\text{inv}}$ regime (left)
 and in the high-$r_{\text{inv}}$ regime (right).}
  \label{fig:sigAUC}
\end{figure*}
The kinematic variables are assigned to each node are
\((k_t, \Delta, z, \psi, m)\), where \(\Delta\) is the angular separation in the Lund plane~\cite{Dreyer:2018nbf}, \(k_t\) is the energy of the branching multiplied by \(\Delta\), \(z\) the energy-sharing fraction, \(\psi\) is the azimuthal angle of the splitting, and \(m\) is the invariant mass of the clustered object. In addition, we include the energy fractions carried by electrons, muons, photons, and hadrons \((f_e, f_\mu, f_\gamma, f_h)\) as proposed in Ref.~\cite{CMS-PAS-EXO-24-029}. By learning directly from this enriched representation of the jet substructure, \texttt{LundNet} captures subtle differences between SVJs and SM jets, enabling a powerful tagging strategy that extends the sensitivity reach of the analysis. All input features are experimentally accessible: the clustering sequence is derived from reconstructed final-state objects, and the splitting variables are kinematic quantities computed from the corresponding pseudo-jets. The particle-type energy fractions \((f_e, f_\mu, f_\gamma, f_h)\) are evaluated from identified particle-flow objects.

Two different \texttt{LundNet} models are trained for the low- and high-\rinv selections, on a mixture of signals and backgrounds. The signal and background samples, after applying the selections, are partitioned into \(80\%\) training, \(10\%\) validation, and \(10\%\) test subsets. In each batch employed during training, the same proportion of signal and background events is maintained. All the signals and backgrounds are represented equally. Further details about the network architecture and training can be found in~\ref{app:train}.

The performance of the trained \texttt{LundNet} models is evaluated by computing the area under the receiver operating characteristic 
curve (AUC). The AUCs over the parameter space of the signal model are summarized in~\figref{fig:sigAUC}, for each of the \texttt{LundNet} models designed for samples in the low- and high-\rinv regimes.

The observed behavior of the classifier performance in terms of the dark hadron mass scale can be understood from the interplay between the energy fractions carried by leptons and photons in the pseudo-jets, and the total number of pseudo-jets in the Lund graph.
First, increasing $\Lambda$ increases the number of pseudo-jets, while increasing \rinv decreases the total number of pseudo-jets. In the low-\rinv scenario, the average fraction of muons over all pseudo-jets in the graph tends to increase with \rinv and decrease with $\Lambda$, see~\ref{app:input}. This trend in the muon energy fraction reproduces the observed AUC performances, suggesting the importance of this feature in discriminating signal from SM jets. 

In the high-\rinv regime, increasing $p_\eta$ lowers \rinv, leading to an increase in the number of pseudo-jets per graph and higher average photon fractions. This enhances the classifier performance, as more photons are present to distinguish signal from background, see~\ref{app:train}.

As detailed in Section~\ref{sec:MC_simulation}, in the low-\rinv scenario we focus on a single benchmark choice for the branching ratios. Extending the analysis to models with different values of $\mathrm{BR}_\ell, \ \mathrm{BR}_q$, and $\mathrm{BR}_\gamma$ would require retraining the GNN to account for these additional signal variations. Further improvements aimed at enhancing robustness against signal-model dependence could be achieved through parametric learning~\cite{Baldi:2016fzo}. However, these developments lie beyond the scope of the present phenomenological study and are therefore not pursued here.

The optimal \texttt{LundNet}  score threshold for jet tagging is determined by scanning the full range of possible values and evaluating the corresponding Significance Improvement Characteristic (SIC) curve~\cite{Gallicchio:2010dq}. The threshold that maximizes this curve is selected as the working point of the tagger. This value is found to be 0.9 in the low- and 0.92 in the high-\rinv regime. Once this value is fixed, events with at least two tagged jets are selected. 

We estimate the expected 95\% confidence level (CL) exclusion limits on the Higgs boson branching ratio into SVJs for different values of the signal model parameters using the modified frequentist \(\mathrm{CL}_{\mathrm{s}}\) method in the asymptotic approximation~\cite{Junk:1999kv,Read:2002hq,Cowan_2011} as implemented in the CMS statistical analysis and combination tool~\cite{CMS:2024onh}. 

Given the rather small number of expected background events in the signal region, systematic uncertainties are expected to be subdominant compared to statistical uncertainties, and are therefore neglected in the discussion below. We have  explicitly verified that the projected sensitivities remain stable when introducing a $1\%$ background normalization uncertainty. A dedicated treatment of tagger systematics is deferred to future work.

\begin{figure*}[t]
  \centering
\includegraphics[width=0.48\linewidth]{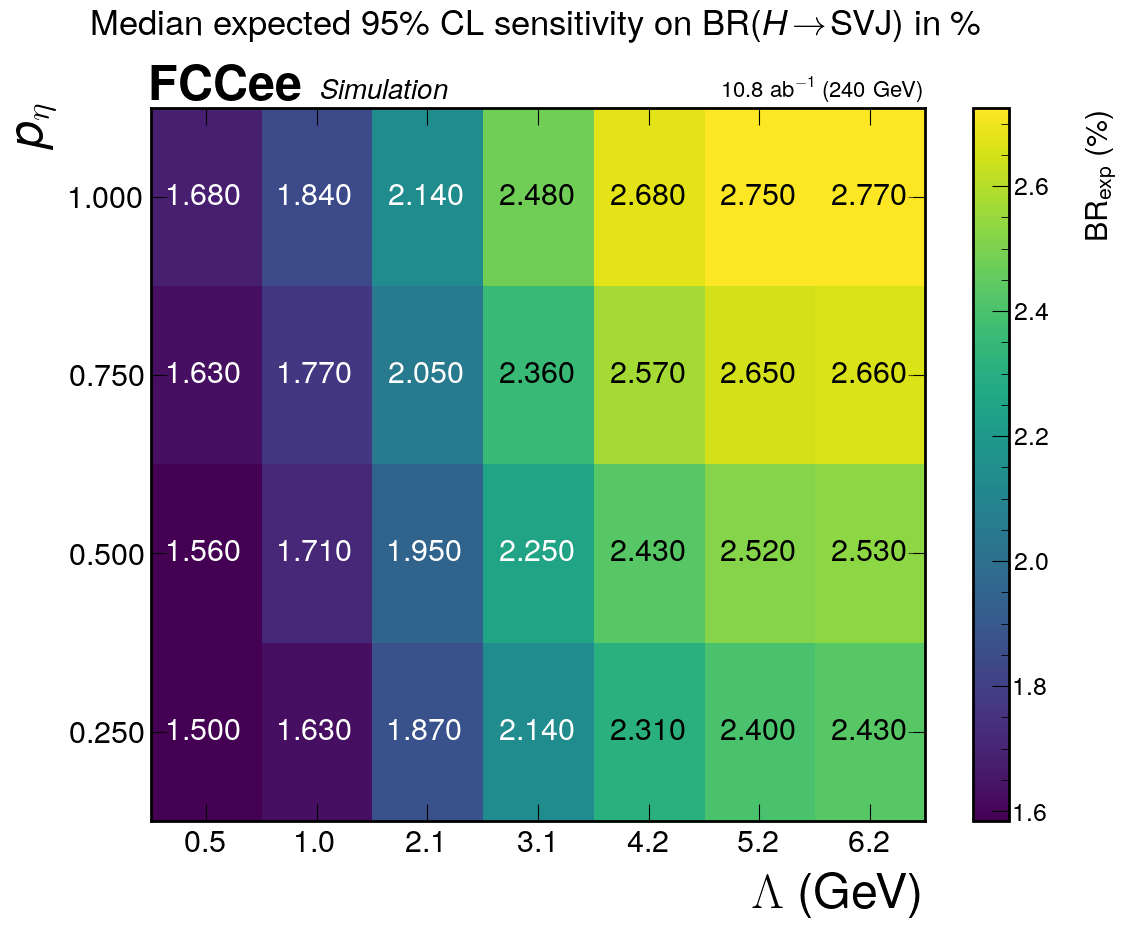}
\includegraphics[width=0.48\linewidth]{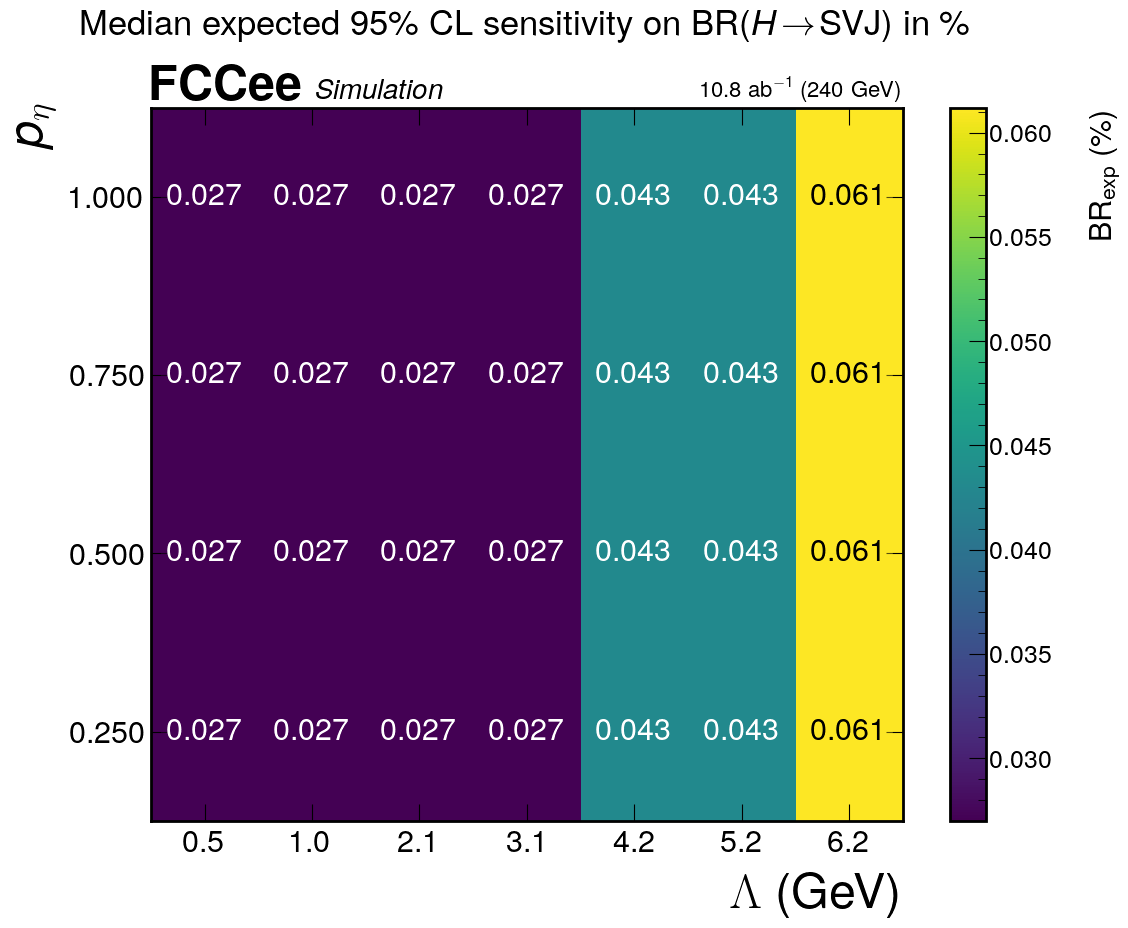}
  \caption{Expected sensitivities in the high $r_{\text{inv}}$ regime without (left) and with (right) the GNN tagger as a function of $\Lambda$ and $p_\eta$ for fixed $p_v=0.5$.}
  \label{fig:limits_high_rinv}
\end{figure*}

\begin{figure*}[t] 
  \centering  
\includegraphics[width=0.48\linewidth]{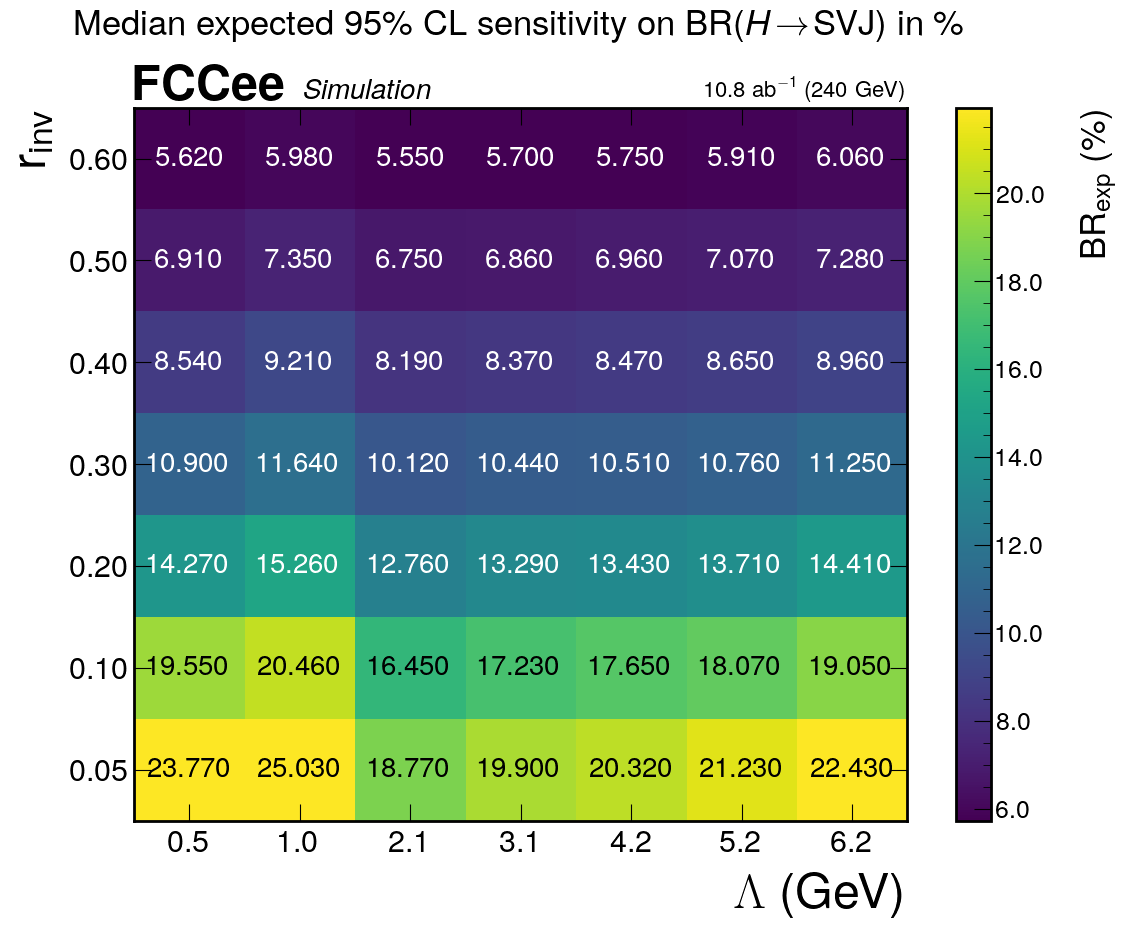}    \includegraphics[width=0.48\linewidth]{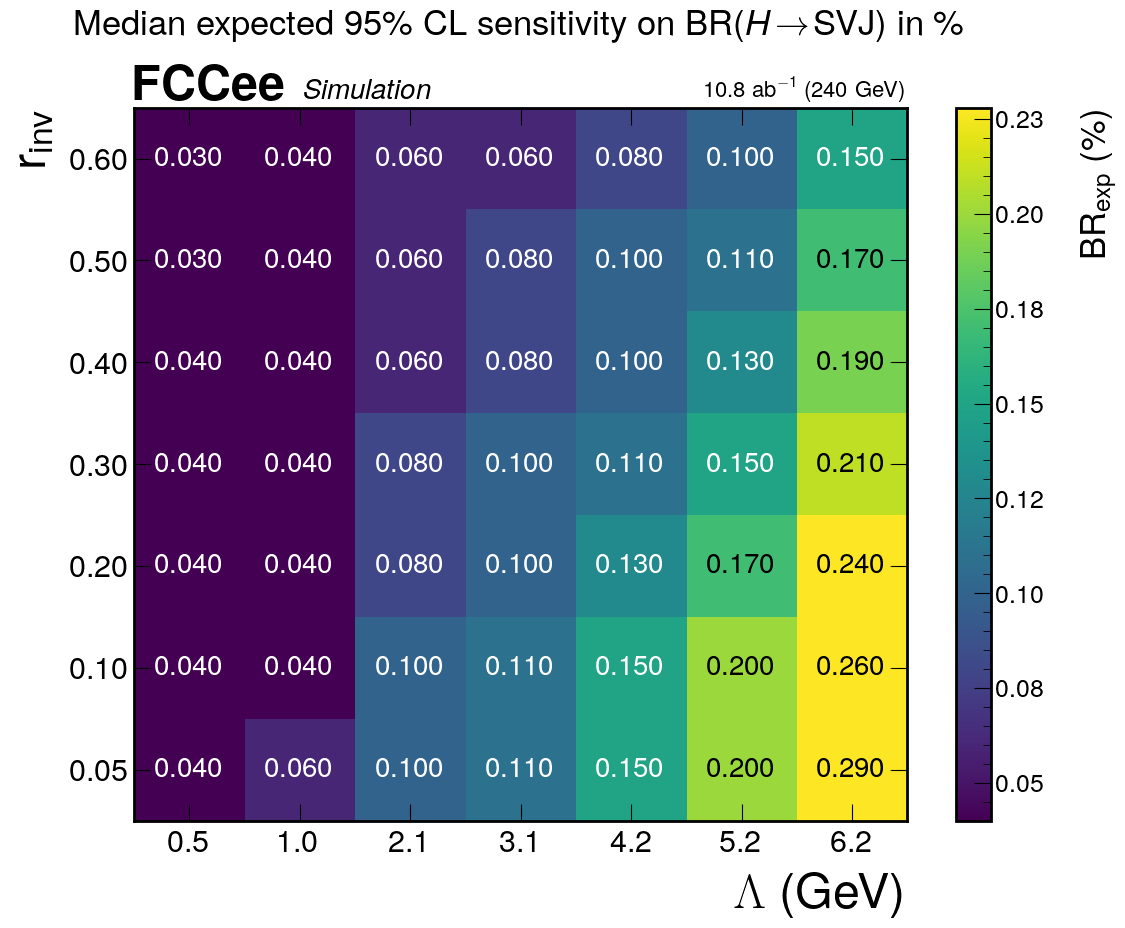}
  \caption{Expected sensitivities in the low $r_{\text{inv}}$ regime without (left) and with (right) the GNN tagger as a function of $(\Lambda,r_{\text{inv}})$ and \(r_{\mathrm{inv}}\).}
  \label{fig:limits-combined}
\end{figure*}

\section{Results}
\label{sec:Res}


The expected sensitivities in the high-\rinv regime with and without a selection based on the \texttt{LundNet}  score, are summarized in~\figref{fig:limits_high_rinv}.
The panels report the $95\%$~CL expected exclusion on $\mathrm{BR}(H \to q_d \bar{q}_d)$ as a function of the relevant mass parameter~$\Lambda$ for representative benchmark choices of $p_v$ and $p_\eta$. A complete set of scans across the full parameter space, including two-dimensional heatmaps in the $(\Lambda, p_v)$ and $(p_\eta, p_v)$ planes, is shown in~\ref{app:sens} for reference. Taken together, these results show that the selection--based strategy achieves strong signal-to-background separation in the high-\rinv regime. Over a broad parameter space, the analysis is very sensitive and is expected to lower the exotic Higgs boson branching ratio limit down to $\mathcal{O}(1\%)$ for the inclusive approach, and $\mathcal{O}(10^{-2}\%)$ for the GNN-based strategy, where the LHC upper limit is around $18\%$~\cite{CMS:2022qva}.

The expected sensitivity in the low-\rinv regime, obtained both with and without the GNN tagger, is shown in~\figref{fig:limits-combined}. The cut-based method alone does not yield optimal sensitivity across the entire parameter space due to the larger background retained after selections. 
This limitation is overcome by applying the \texttt{LundNet} tagger, which enables sensitivity to $\mathrm{BR}(H \to q_d \bar{q}_d)$ down to the permille level in the full parameter space considered.

By using eq.~\eqref{eq:HiggsPortal_width}, it is possible to translate these sensitivities into expected upper limits on the coupling $y$. Across the full parameter space -- including both the low- and high-\rinv regimes -- the median expected 95\% CL upper limit on the coupling $y$ ranges from $4.1 \times 10^{-4}$ to $1.3 \times 10^{-3}$.

\section{Conclusions}
\label{sec:Conc}
In this work, we developed and tested a dedicated search for SVJs at the FCC-ee operating at \(\sqrt{s} = \SI{240}{GeV}\), with production mediated by the Higgs boson and leveraging the clean \(Z \to \ell^+ \ell^-\) recoil topology. We constructed a benchmark program that (i) maps the effective invisible fraction \rinv to the underlying hadronization parameters \((p_v, p_\eta, \Lambda)\) through simulation, and (ii) scans both a high-\rinv scenario, where \rinv is a derived quantity, and a low-\rinv scenario, where \rinv is taken as an explicit input. This provides a coherent way to navigate between regimes dominated by genuine \(E_{\mathrm{miss}}\) and those with predominantly visible final states.

In the high-\rinv regime, we designed a selection strategy based on an \(E_{\mathrm{miss}}\) window and an azimuthal separation requirement \(\Delta\phi\), which shows sensitivity at the percent level for \(\mathrm{BR}(H \to q_d \bar{q}_d)\). Moreover, applying the GNN tagger on top of this selection further extends the reach: the sensitivity to \(\mathrm{BR}(H \to q_d \bar{q}_d)\) improves by nearly three orders of magnitude, substantially enhancing the overall discovery potential.
In the low-\rinv regime, the invisible energy is intrinsically smaller: very tight \(\Delta\phi\) requirements become overly punitive for the signal, and purely cut-based selections generally fall short of the high-sensitivity target. To address this specifically in the low-\rinv regime, we show that the GNN tagger is crucial, allowing constraints on \(\mathrm{BR}(H \to q_d \bar{q}_d)\) at the level of $\mathcal{O}(10^{-2}~\%)$.

 These results can be directly translated into constraints on the Higgs portal coupling $y$. Over the full parameter space, spanning both the low- and high-\rinv regimes, we obtain expected 95\% CL upper limits on $y$ at the $\mathcal{O}(10^{-3})$ level, demonstrating the impressive sensitivity of the proposed analysis.

 Overall, the FCC-ee shows strong potential to probe Higgs-portal dark sectors leading to SVJ signatures: optimized selections suffice for discovery-level sensitivity in the high-\rinv regime, while modern jet-substructure tagging specifically enhances sensitivity at low \rinv, pushing SVJ dynamics well within experimental reach. Furthermore, the simplified modeling strategy adopted here can be readily extended to hadron colliders such as the FCChh and LHC, where complementary searches would provide an even broader test of the parameter space for SVJs, which is currently under investigation.

\begin{acknowledgements}
The authors would like to warmly thank Juraj Smiesko for the valuable discussions and support with the FCC framework. We also thank the the FCC collaboration for endorsing this work, and particularly Juliette Alimena for useful comments and suggestions on the paper draft. 
C. Cazzaniga, A. de Cosa and R. Seidita are supported by the Swiss National Science Foundation (SNSF) under the SNSF Eccellenza program ($\mathrm{PCEFP2}\_\mathrm{186878}$). F.~Kahlhoefer acknowledges funding from the Deutsche Forschungsgemeinschaft
(DFG) through Grant No. 396021762 -- TRR 257.

\end{acknowledgements}
 
\appendix

\clearpage
\onecolumn
\clearpage
\twocolumn

\section{Plots of invisible fraction}\label{app:plot_rinv}
\begin{figure}[H]
  \centering

  \begin{subfigure}{0.9\linewidth}
    \includegraphics[width=\linewidth]{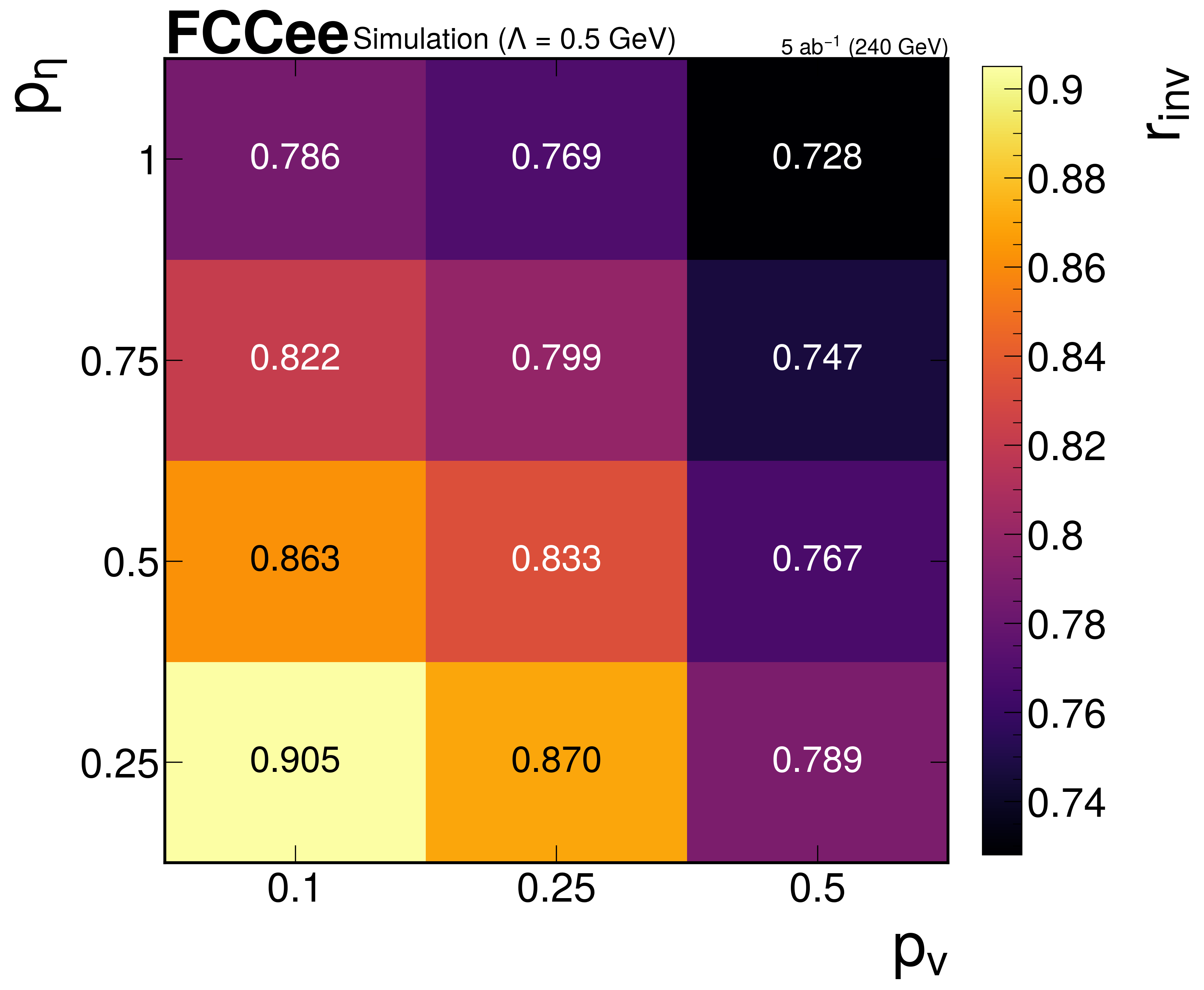}
    \caption{$\Lambda=\SI{0.521}{GeV}$}
  \end{subfigure}\hfill
  \begin{subfigure}{0.9\linewidth}
    \includegraphics[width=\linewidth]{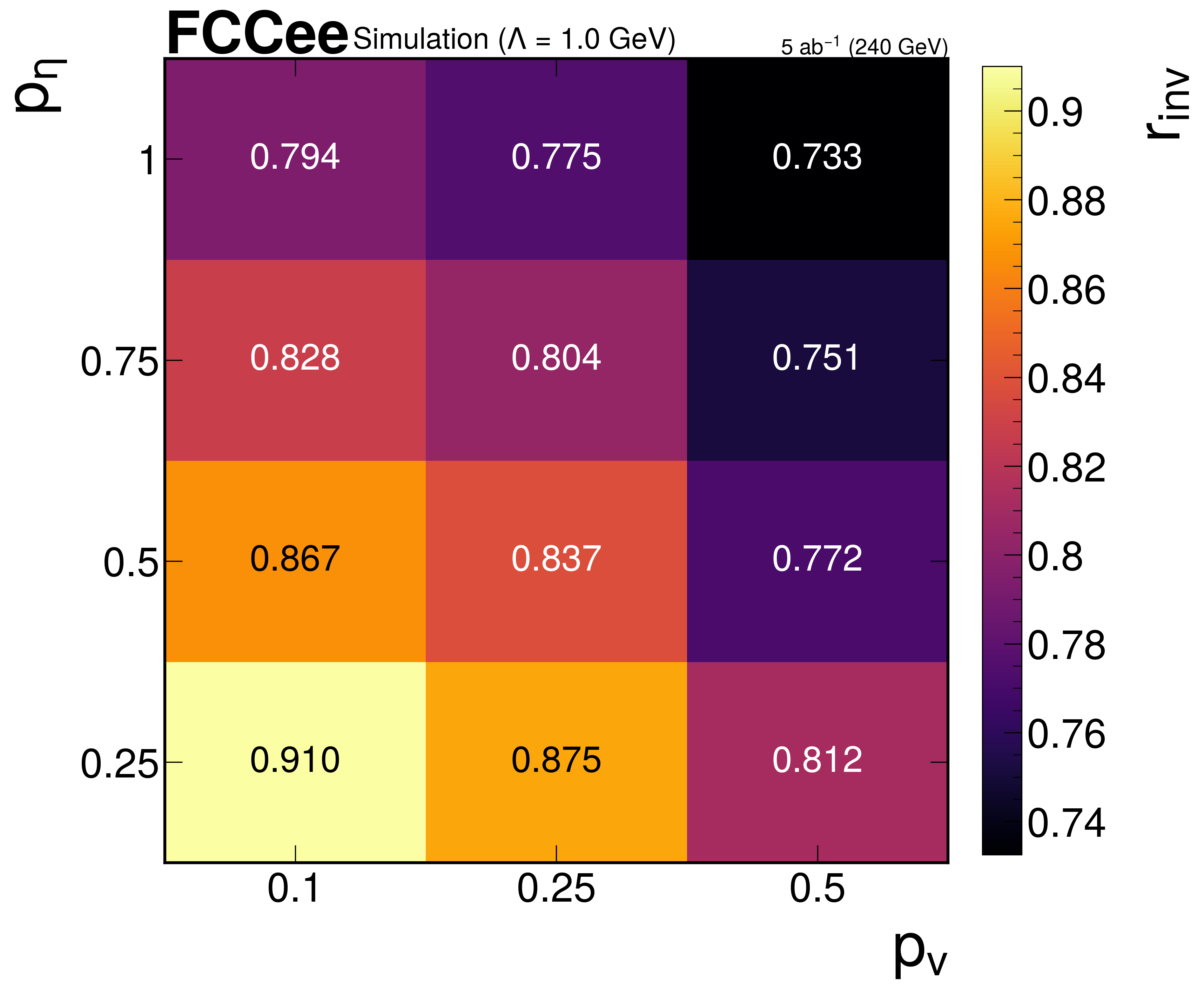}
    \caption{$\Lambda=\SI{1.042}{GeV}$}
  \end{subfigure}

  \medskip

  \begin{subfigure}{0.9\linewidth}
    \includegraphics[width=\linewidth]{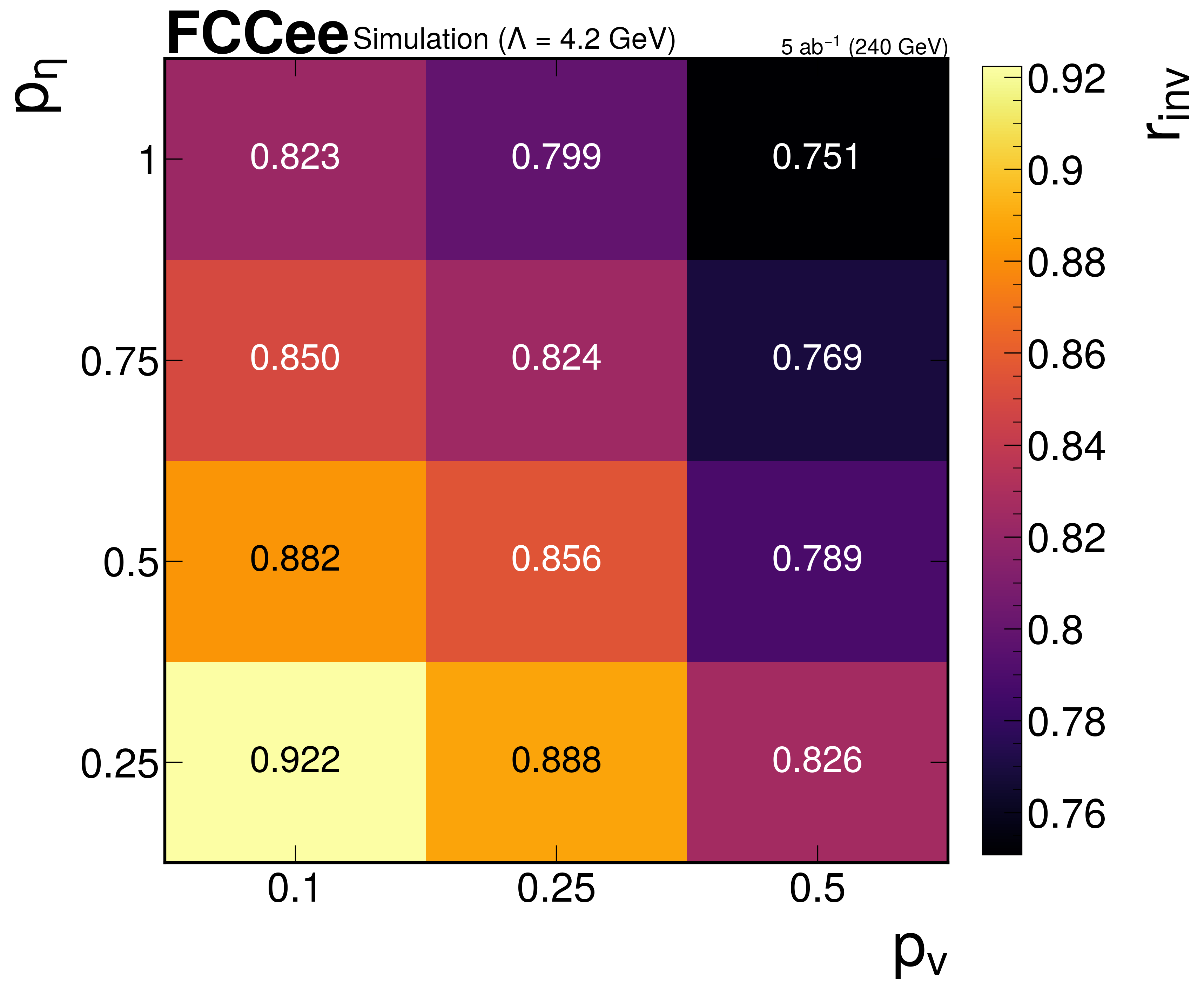}
    \caption{$\Lambda=\SI{4.167}{GeV}$}
  \end{subfigure}\hfill

  \medskip

  \caption{$r_{\mathrm{inv}}$ values obtained with simulation data changing $p_v$ and $p_\eta$ for different $\Lambda$ values.}
  \label{fig:rinv}
\end{figure}

Here we study how \(r_{\mathrm{inv}}\) is connected to the parameters $p_v$ and $p_\eta$ in the model employed for the high-\(r_{\mathrm{inv}}\) scenario. For this study we employ simulations since there is no strict analytical formulation relating the parameters considered. In this study, for each signal point we simulate a sample, and compute \(r_{\mathrm{inv}}\) event-by-event counting the number of stable and unstable dark hadrons. In Figure~\ref{fig:rinv}, we plot the average over the full MC sample employed. We show that \(r_{\mathrm{inv}}\) tends to decrease with $p_v$ and $p_\eta$. 

\section{Signal Efficiencies}\label{app:sigeff}

\begin{figure}[t]  
  \centering
    \includegraphics[width=0.99\linewidth]{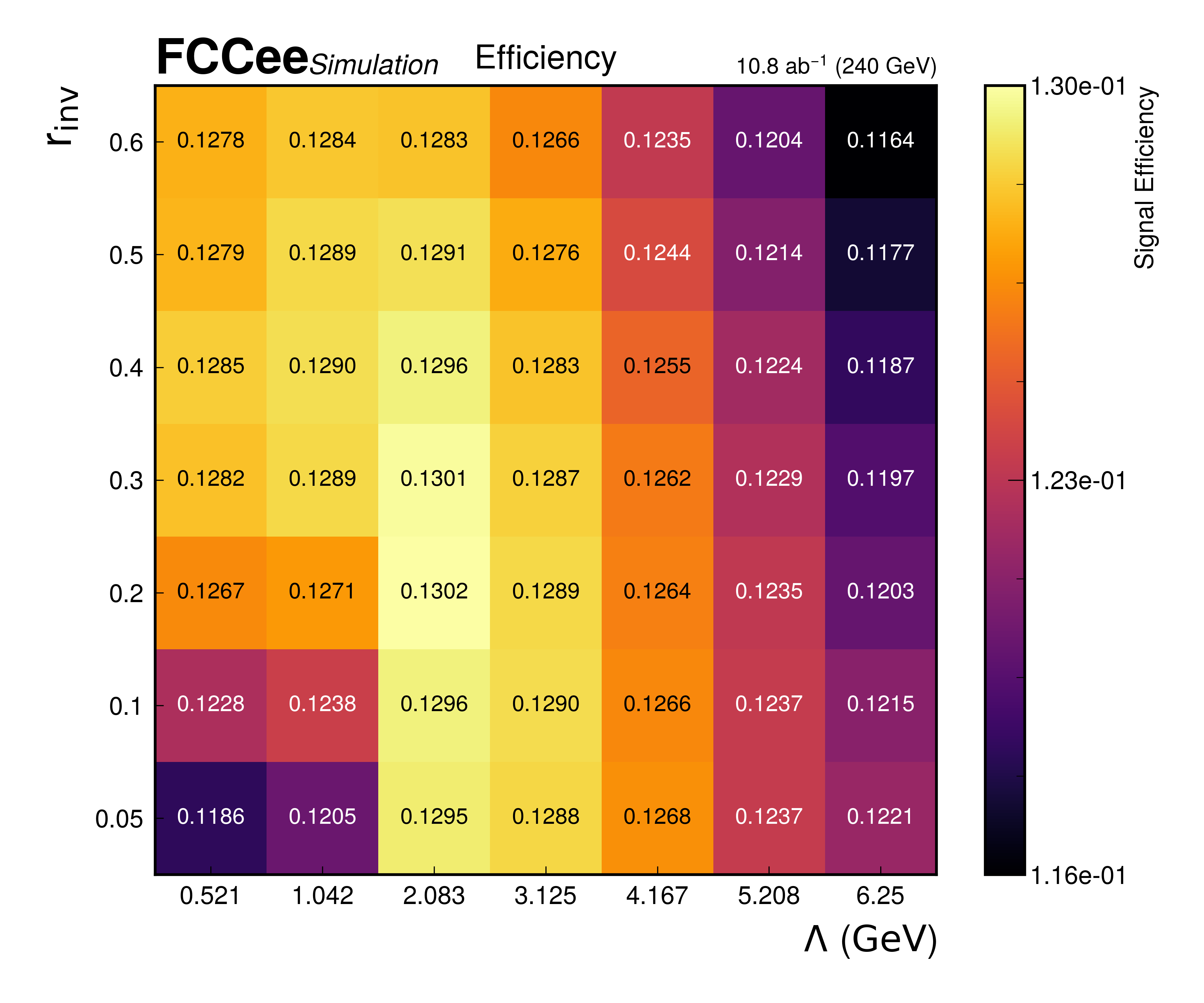}
    \caption{Signal efficiency in $(\Lambda, r_{\text{inv}})$ for the low-$r_{\text{inv}}$ scenario.}
 \label{fig:signal_eff_lowrinv} 
\end{figure}

\begin{figure}[H]  
  \centering
  \begin{subfigure}{0.95\linewidth}
    \centering
    \includegraphics[width=\linewidth]{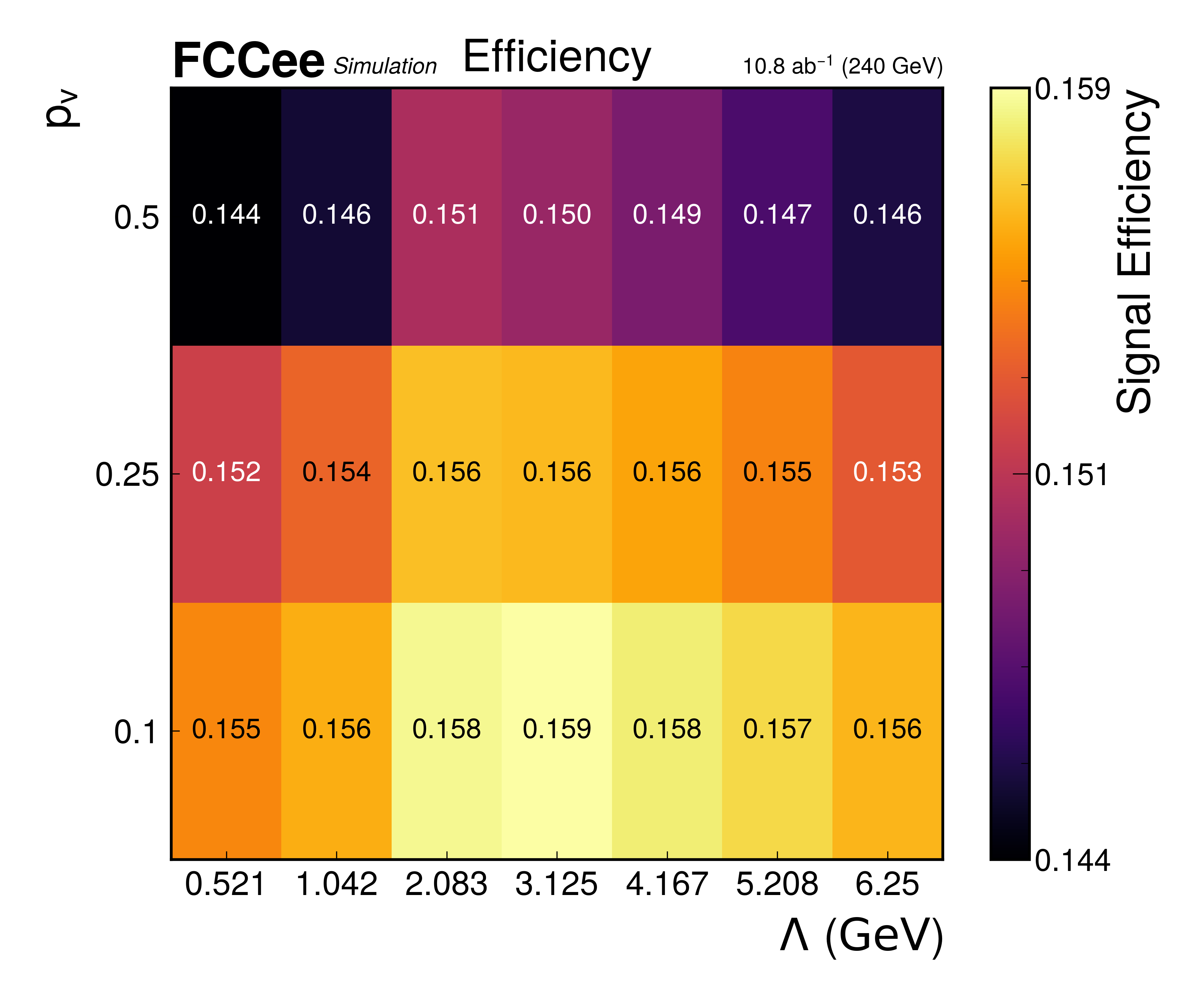}
    \caption{Signal efficiency in $(\Lambda, p_{v})$}
  \end{subfigure}\hfill
  \begin{subfigure}{0.95\linewidth}
    \centering
    \includegraphics[width=\linewidth]{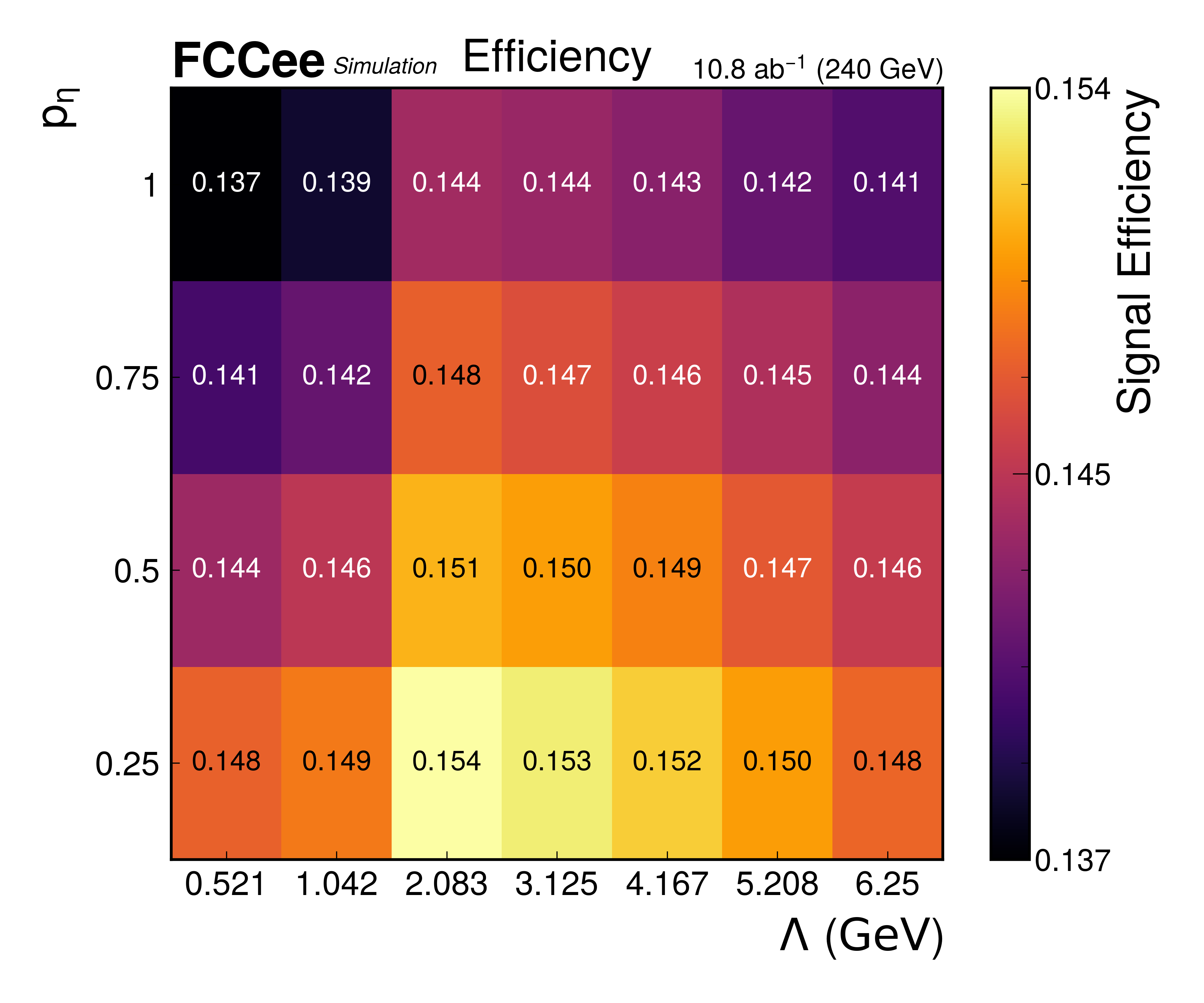}
    \caption{Signal efficiency in $(\Lambda, p_{\eta})$}
  \end{subfigure}

\medskip

  \begin{subfigure}{0.95\linewidth}
    \centering
    \includegraphics[width=\linewidth]{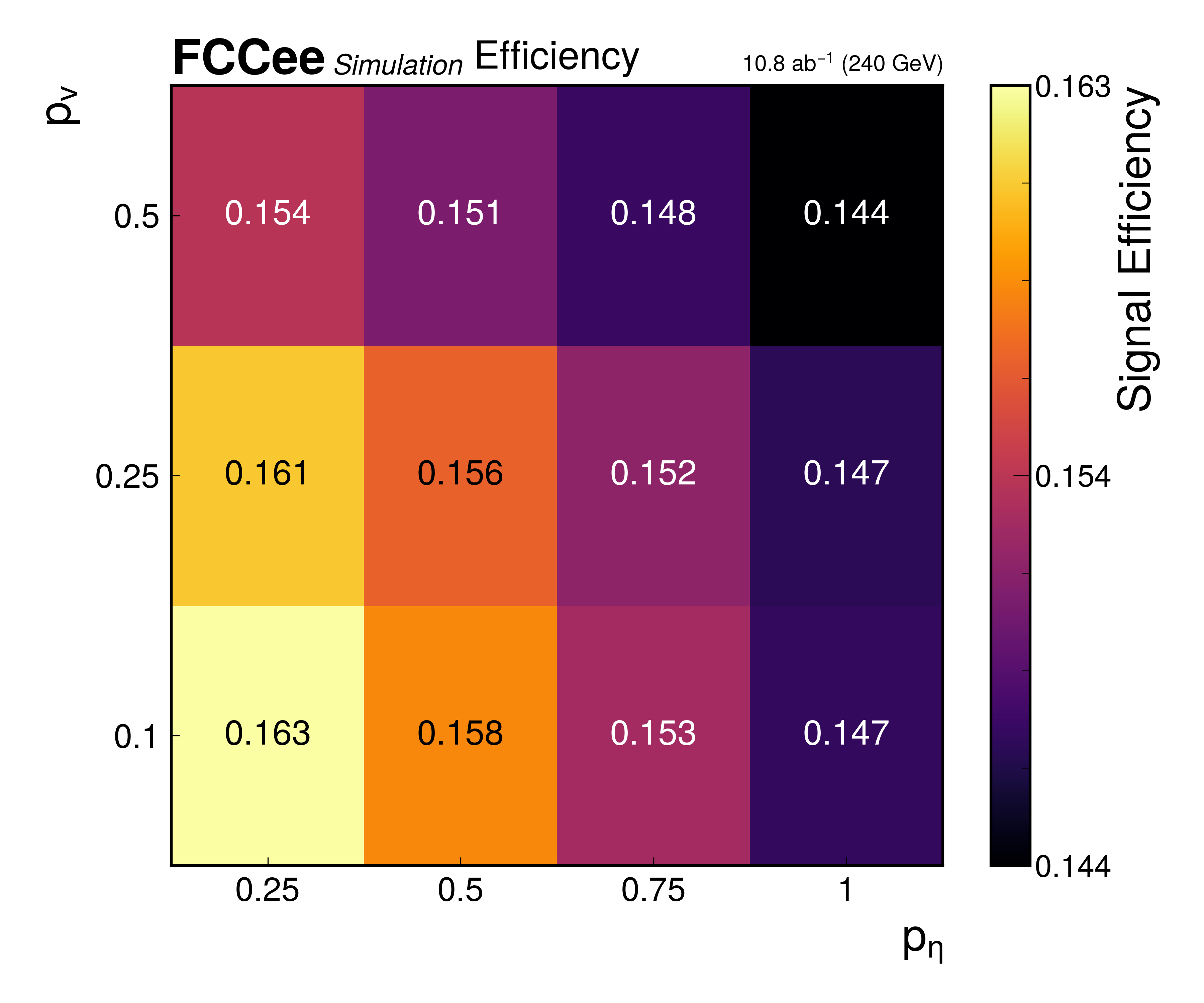}
    \caption{Signal efficiency in $(p_{\eta}, p_{v})$}
  \end{subfigure}
\caption{Signal efficiencies in the parameter space for the high-$r_{\text{inv}}$ scenario.}
 \label{fig:signal_eff_highrinv} 
\end{figure}

For the selections in the low and high $r_{\mathrm{inv}}$ regimes, defined in Tables~\ref{tab:selection_cuts}, the final signal selection efficiency is studied in the parameter space of the models. 
In the low-$r_{\mathrm{inv}}$ regime, the signal efficiency is shown in the top panel of Figure~\ref{fig:signal_eff_lowrinv} as a function of $\Lambda$ and $r_{\text{inv}}$. For the high-$r_{\mathrm{inv}}$ regime, Figure~\ref{fig:signal_eff_highrinv} shows the signal efficiency in the parameter space of the model for different combinations of $p_v$, $p_\eta$ and $\Lambda$.

\section{LundNet architecture and further performances}
\label{app:train}

 For completeness, we summarize here the architecture and training setup used for the \texttt{LundNet} classifier.

\begin{itemize}
  \item \textbf{Classifier:} \texttt{LundNet}, implemented as a single \texttt{EdgeConv} block with mean aggregation, followed by global mean pooling.  
  \item \textbf{EdgeConv block:} three linear layers  
  \(2d \;\to\; 32 \;\to\; 32 \;\to\; 64\), each with batch normalization and ReLU.  
  \item \textbf{Dense head:} \(64 \;\to\; 64 \;\to\; 32 \;\to\; 1\), with intermediate batch normalization, ReLU activations, dropout (\(p=0.2\)); final sigmoid output.  
  \item \textbf{Training:} Adam optimizer, learning rate \(\eta=10^{-3}\), no weight decay, up to 500 epochs. LR reduced by 0.5 on plateau.  
  \item \textbf{Early stopping:} patience 20, minimum 80 epochs, tolerance \(\Delta=1.5\times10^{-3}\).  
  \item \textbf{Batch size:} 64 (training), 128 (evaluation).  
  \item \textbf{Loss:} binary cross-entropy with class weighting.  
\end{itemize}

\begin{figure}[t]
  \centering
  \begin{subfigure}{0.48\textwidth}
    \centering
    \includegraphics[width=\linewidth]{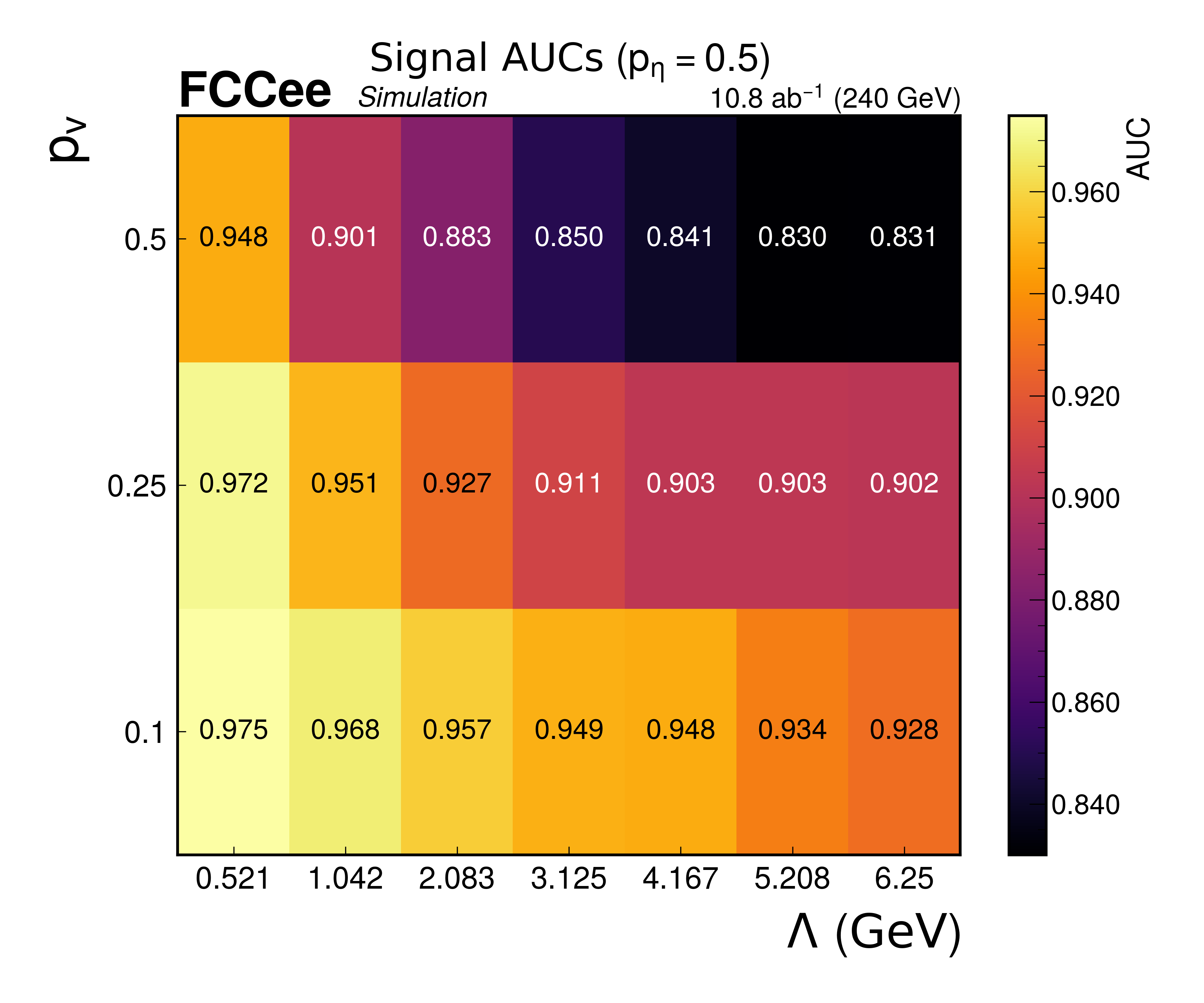}
    \caption{(a) AUCs for $(\Lambda, p_v,p_{\eta}=0.5)$}
    \label{fig:auclambdapv}
  \end{subfigure}\hfill
  \begin{subfigure}{0.48\textwidth}
    \centering
    \includegraphics[width=\linewidth]{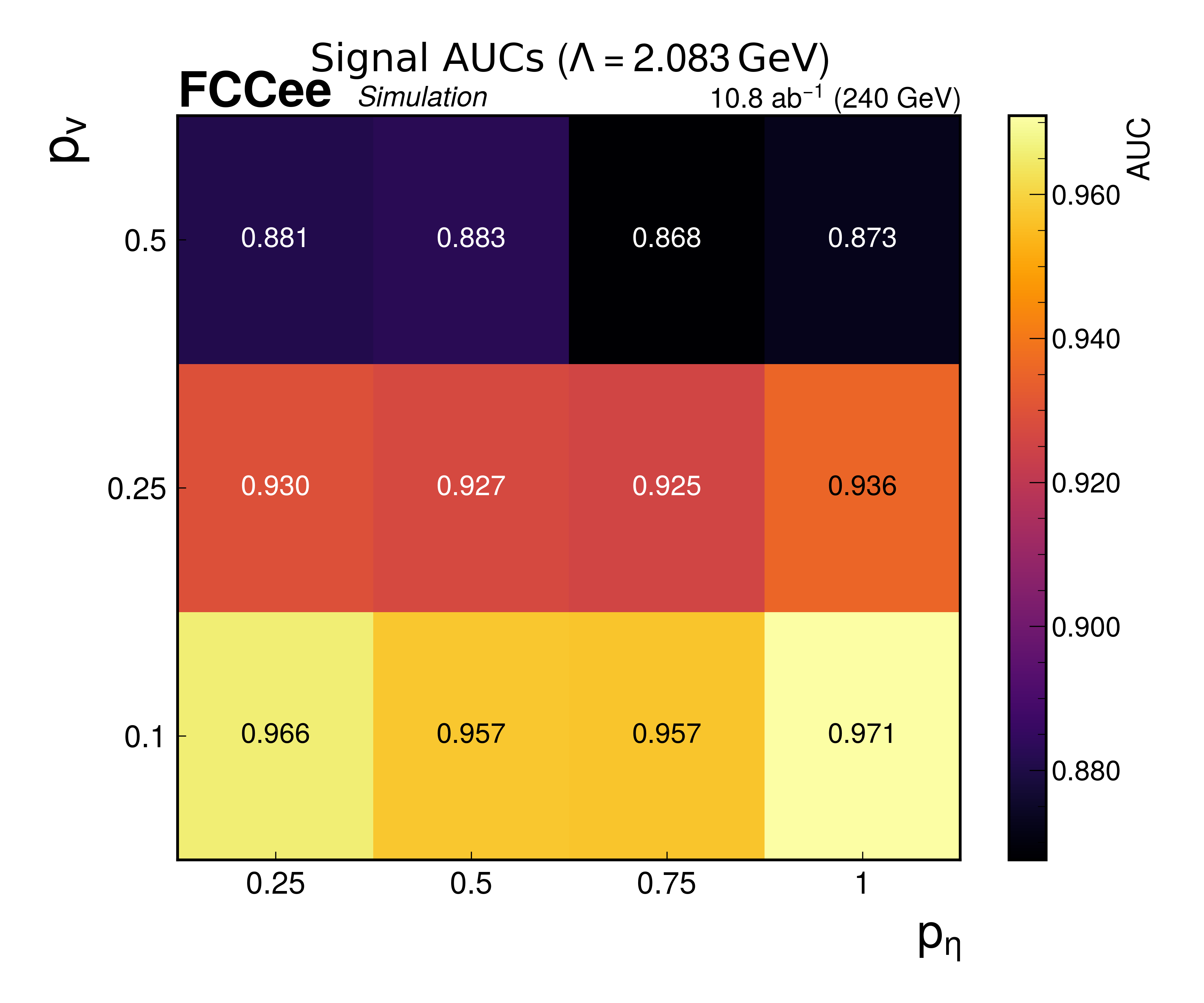}
    \caption{(b) AUCs for $(p_{\eta}, p_v, \Lambda=\SI{2.083}{GeV} )$}
    \label{fig:aucpetapv}
  \end{subfigure}
  \caption{High-$r_{\text{inv}}$ AUCs for the remaining parameter space}
  \label{fig:auchighrest}
\end{figure}

 The performance of the GNN in the remaining part of the parameter space for high-$r_{\mathrm{inv}}$ not reported in the main text is shown in Figure~\ref{fig:auchighrest}.

\clearpage

\section{Averaged input features to the GNN}
\label{app:input}

\begin{figure}[t]
  \centering
  \begin{subfigure}{0.48\textwidth}
    \centering
    \includegraphics[width=\linewidth]{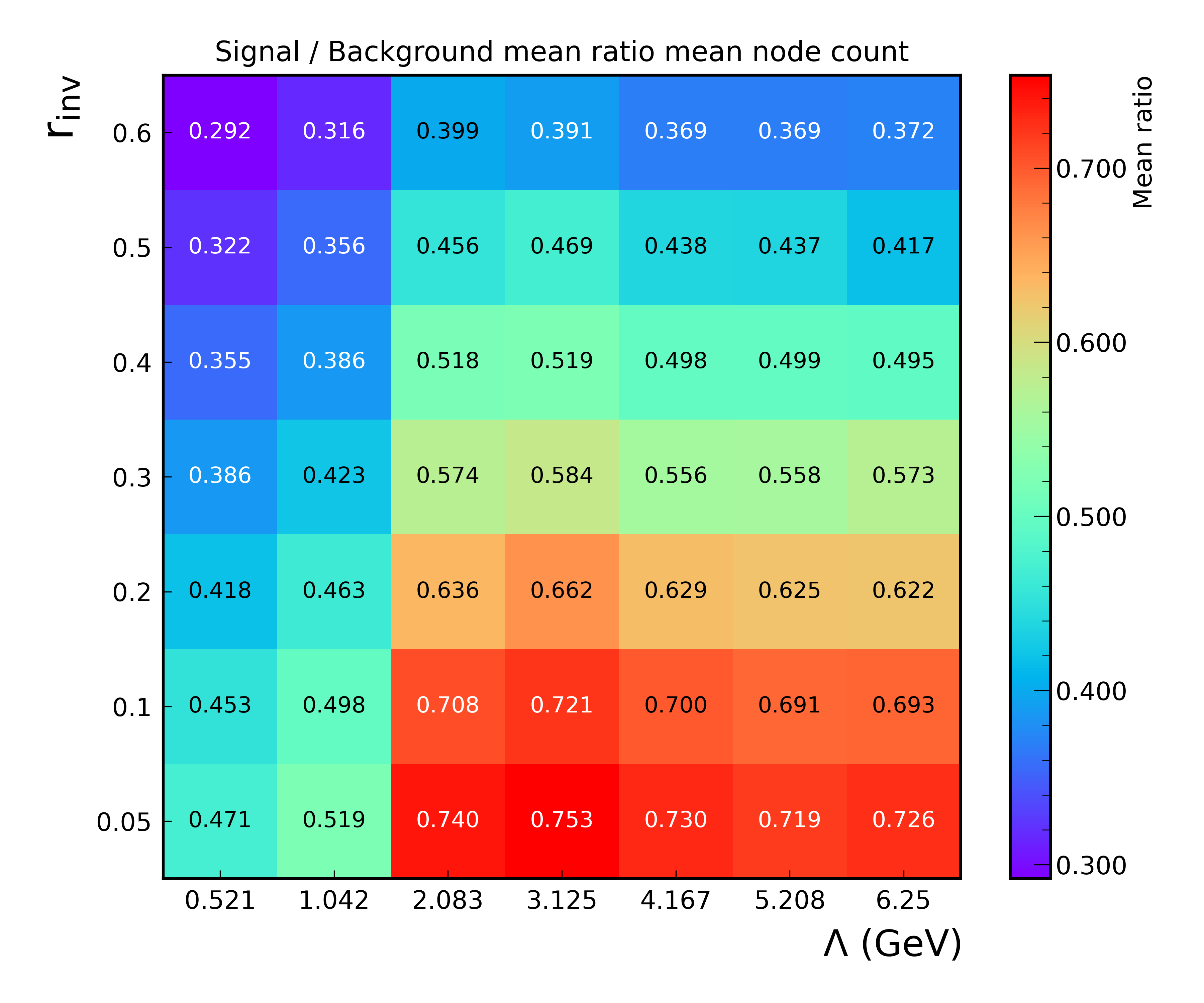}
    \caption{Ratio between the average number of nodes per input graph for different signals and total background.}
    \label{fig:mean_nodes}
  \end{subfigure}\hfill
  \begin{subfigure}{0.48\textwidth}
    \centering
    \includegraphics[width=\linewidth]{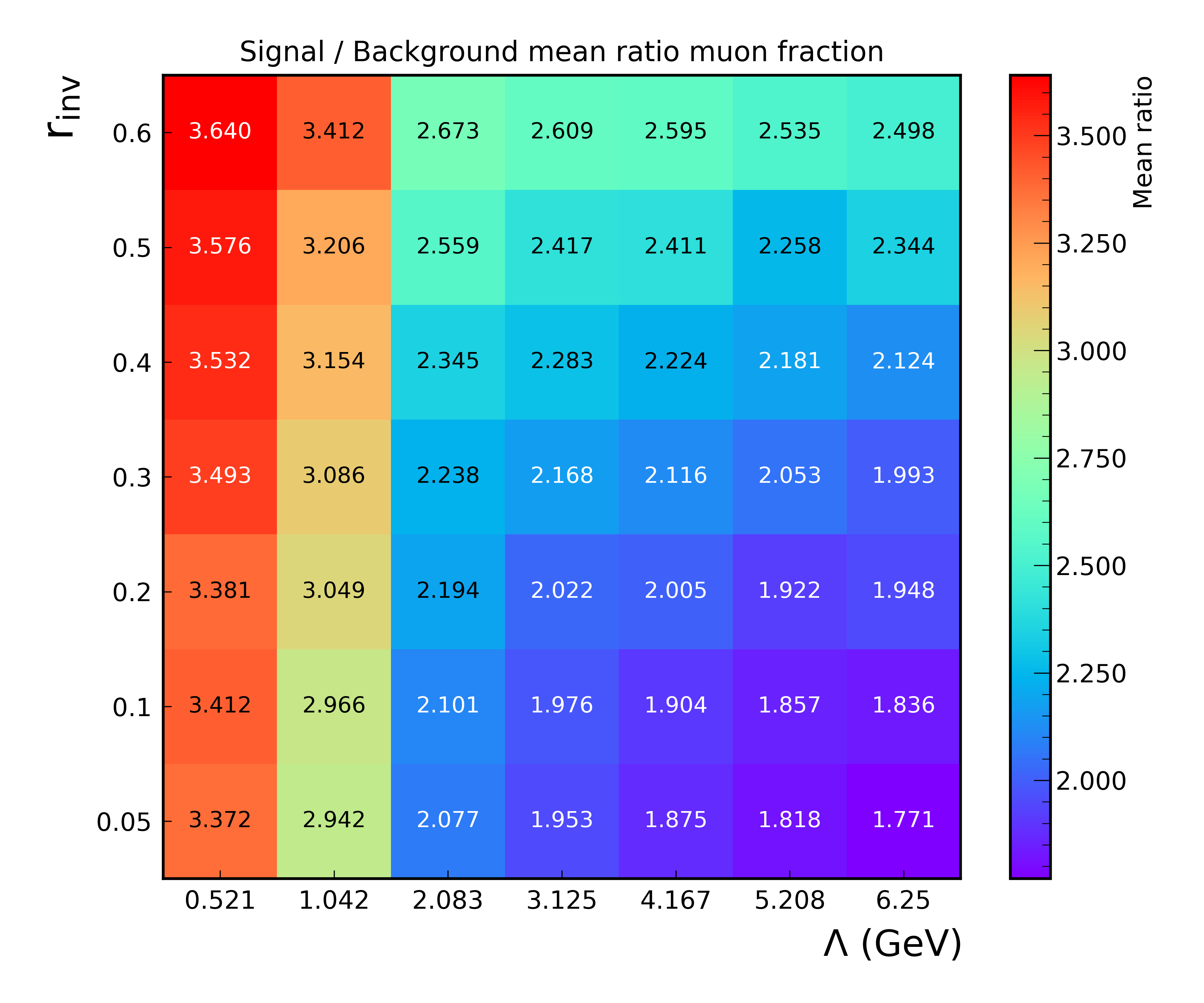}
    \caption{Ratio between the average muon energy fraction per input graph for different signals and total background.}
    \label{fig:mean_muons}
  \end{subfigure}
  \caption{Averaged input features for all nodes of all graphs in the low-$r_{\text{inv}}$ scenario. }
  \label{fig:averages_low_rinv}
\end{figure}

In this section we show how some of the input features attached to each node of the graph change in the parameter space of the signal. In specific, we compute for each signal and background graph the average of the node feature over all the nodes. Then, this feature is averaged over all the graphs employed for the training of the GNN. In Figures~\ref{fig:mean_nodes} and~\ref{fig:mean_muons}, the averaged number of nodes and muon energy fraction are shown. These features are relevant for explaining the trends in the performance plots of the GNN, shown in Figure~\ref{fig:sigAUC}.

\clearpage

\begin{figure}[t]
  \centering
  \begin{subfigure}{0.48\textwidth}
    \centering
    \includegraphics[width=\linewidth]{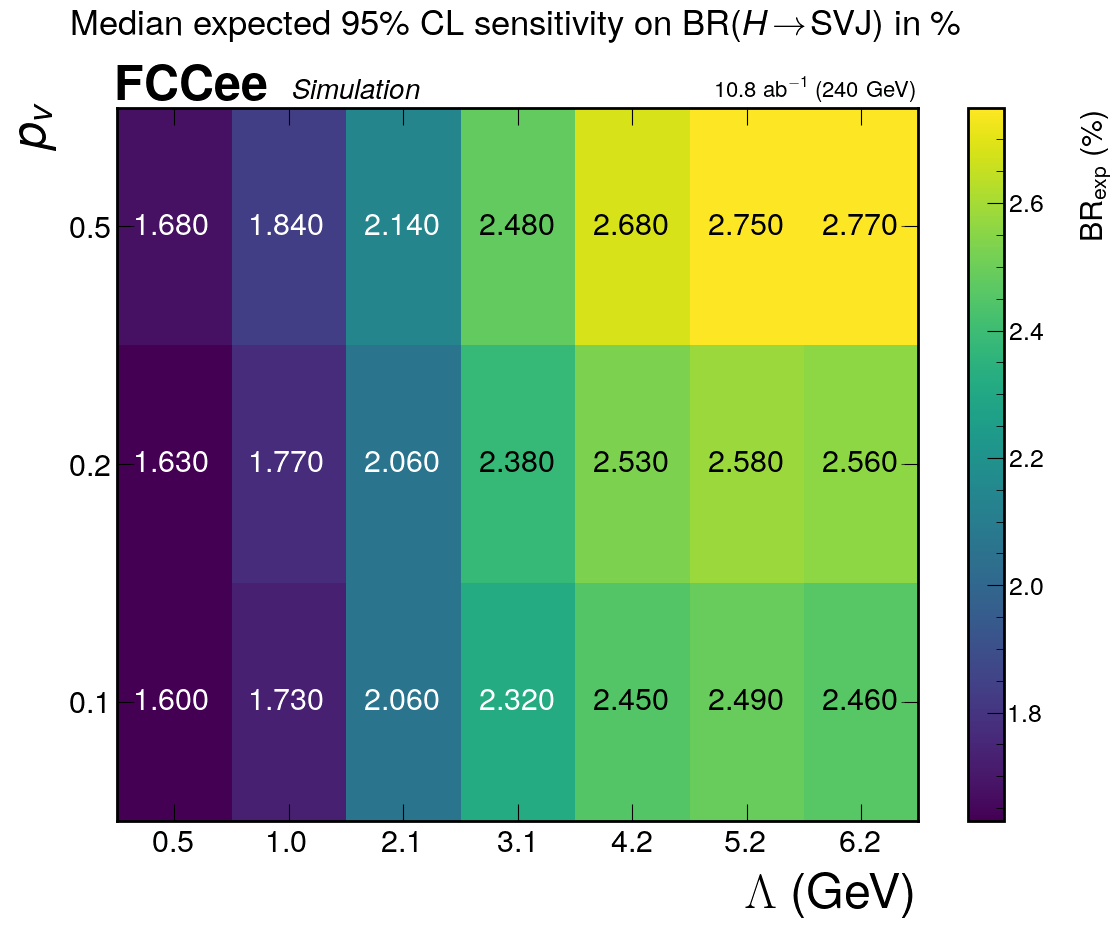}
    \caption{(a) Sensitivity vs.\ $\Lambda$ at fixed $p_\eta=0.5$.}
    \label{fig:limits_high_rinv1}
  \end{subfigure}\hfill
  \begin{subfigure}{0.48\textwidth}
    \centering
    \includegraphics[width=\linewidth]{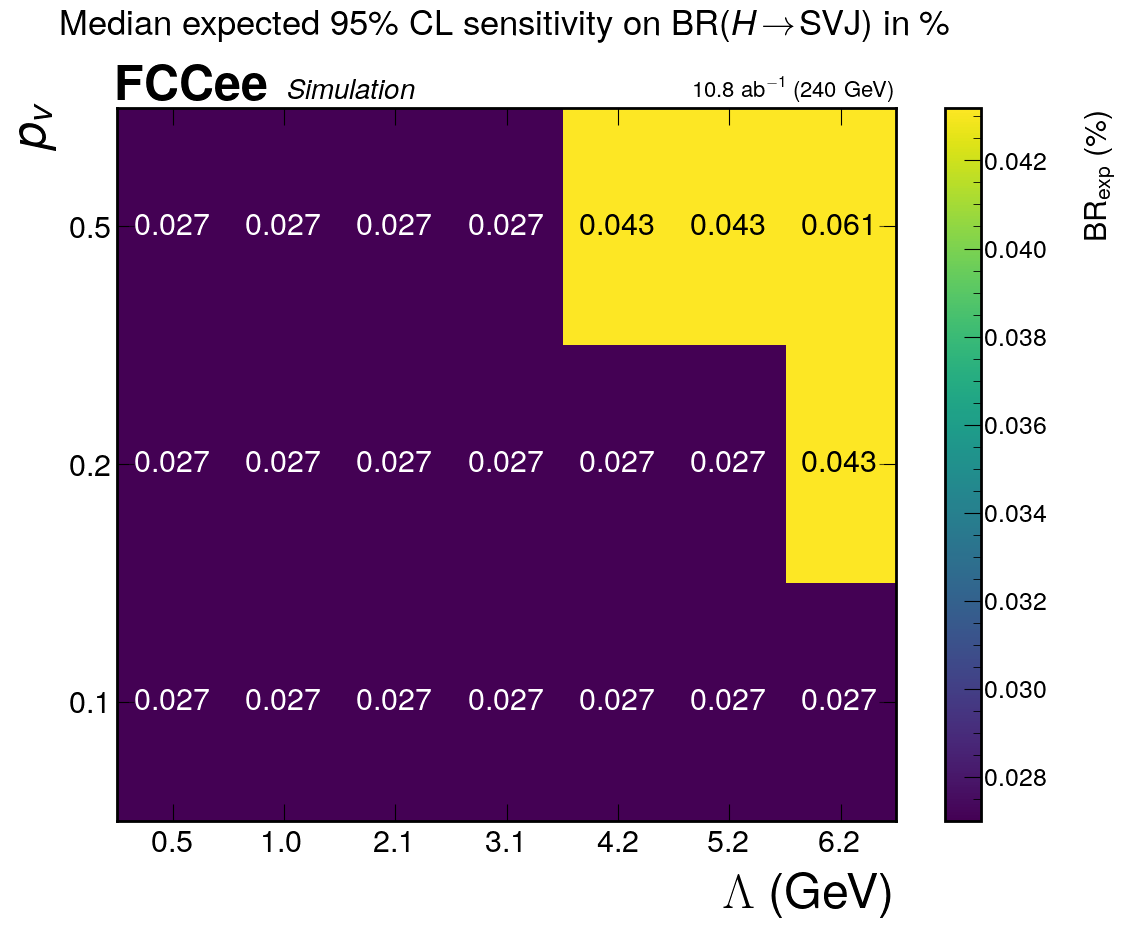}
    \caption{(b) Sensitivity vs.\ $\Lambda$ at fixed $p_\eta=0.5$ with GNN tagger}
    \label{fig:limits_other1}
  \end{subfigure}
  \caption{Sensitivity limits without (a) and with (b) GNN tagger}
  \label{fig:limits-peta-slices1}
\end{figure}

\section{Sensitivity plots over the full high \texorpdfstring{$r_{\mathrm{inv}}$}{r_inv} parameter space}
\label{app:sens}

\begin{figure}[t]
  \centering
  \begin{subfigure}{0.48\textwidth}
    \centering
    \includegraphics[width=\linewidth]{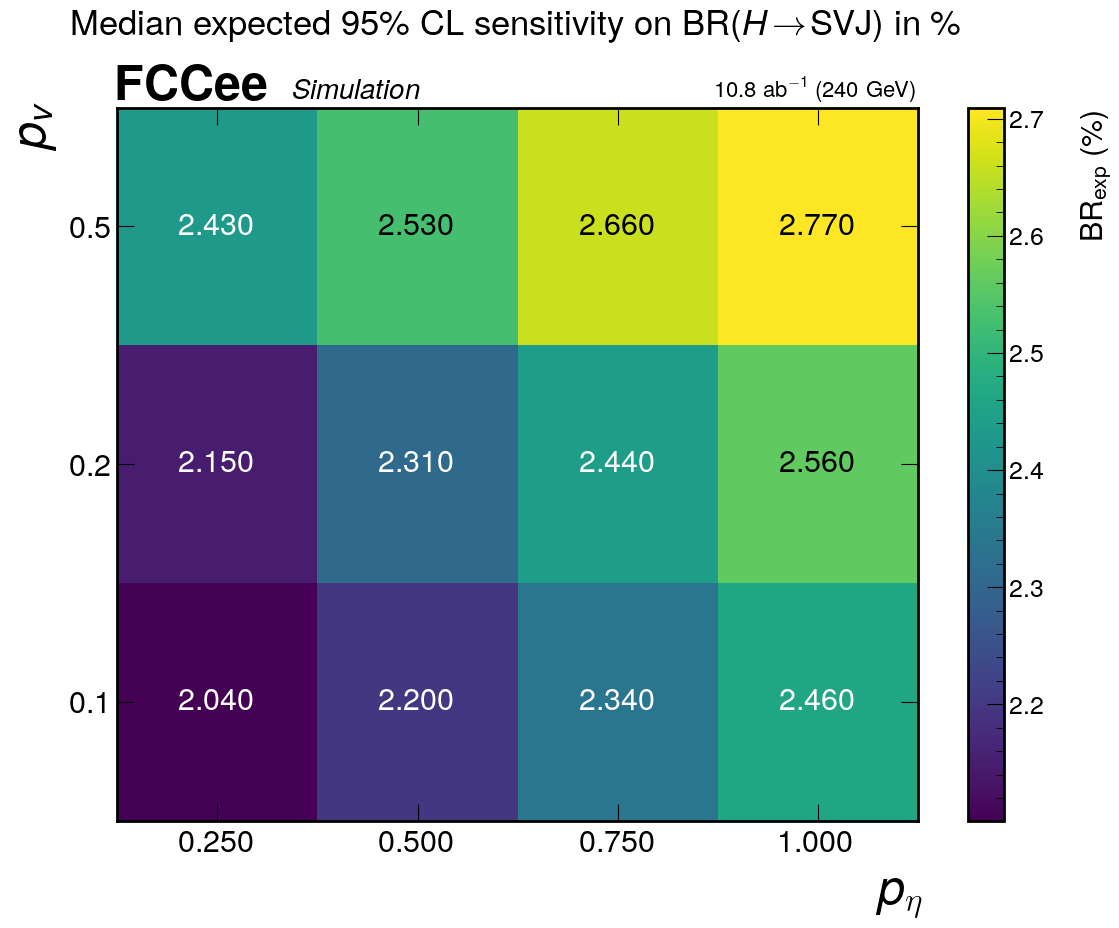}
    \caption{(a) Sensitivity vs.\ $p_\eta$ at fixed $\Lambda=\SI{2.083}{GeV}$.}
    \label{fig:limits_high_rinv2}
  \end{subfigure}\hfill
  \begin{subfigure}{0.48\textwidth}
    \centering
    \includegraphics[width=\linewidth]{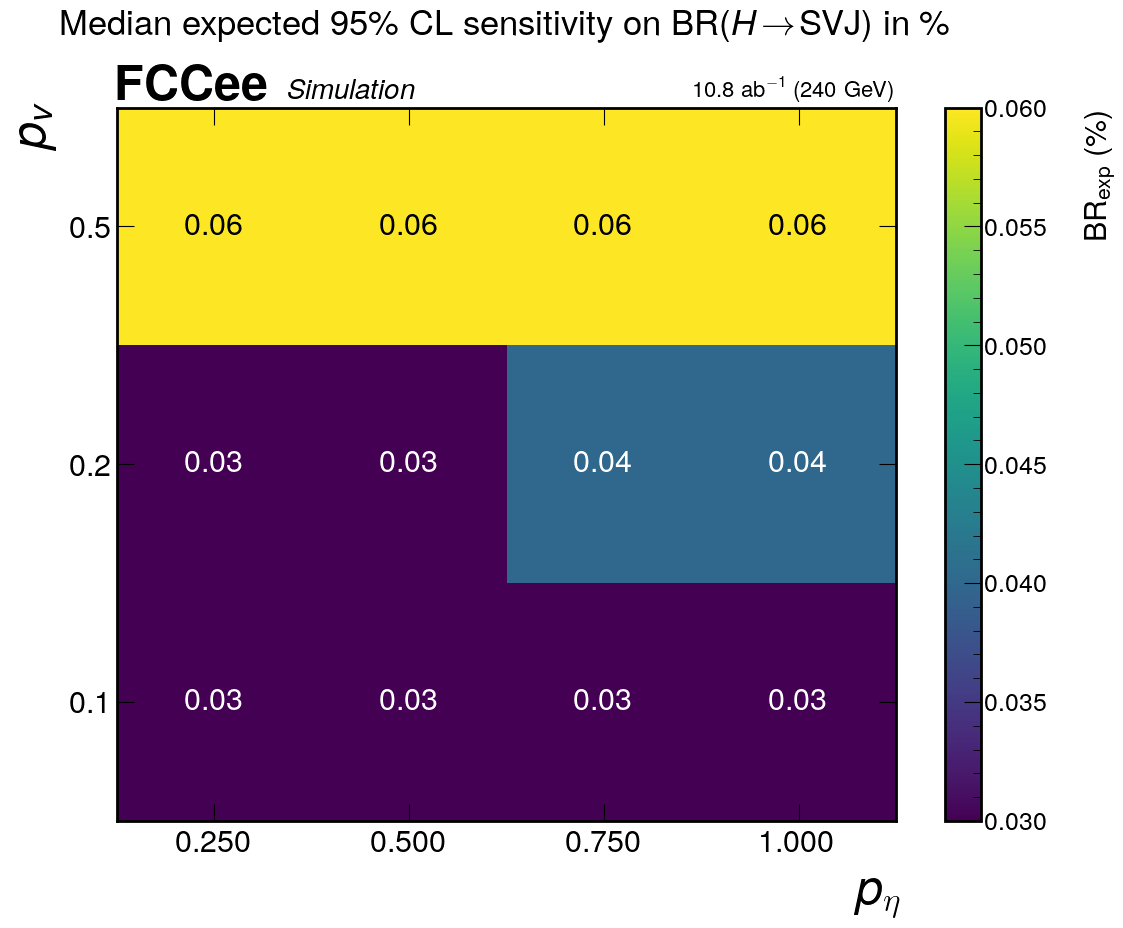}
    \caption{(b) Sensitivity vs.\ $p_\eta$ at fixed $\Lambda=\SI{2.083}{GeV}$ with GNN tagger}
    \label{fig:limits_other2}
  \end{subfigure}
  \caption{Sensitivity limits without (a) and with (b) GNN tagger}
  \label{fig:limits-peta-slices2}
\end{figure}

In Figures~\ref{fig:limits-peta-slices1} and~\ref{fig:limits-peta-slices2}, we show how the sensitivity of the proposed strategy changes for different parameter combinations not shown in the main text for the high-$r_{\text{inv}}$ regime when applying the GNN.

\clearpage
\bibliographystyle{spphys} 
\bibliography{bibl}   

\end{document}